\documentclass[10pt,a4paper]{article}

\usepackage{caption}
\usepackage{subcaption}
\usepackage{graphicx}
\graphicspath{{figures/}}

\usepackage{amssymb}
\usepackage{amsmath} 

\usepackage{algorithmic}
\usepackage{array}
\usepackage{lipsum}
\usepackage{hyperref}
\usepackage[utf8]{inputenc} 
\usepackage[T1]{fontenc}    
\usepackage{hyperref}       
\hypersetup{
   citecolor=blue,
    colorlinks=true,
    linkcolor=red,
    filecolor=magenta,      
    urlcolor=blue,
linktocpage}
\usepackage{url}            
\usepackage{booktabs}       
\usepackage{amsfonts}       
\usepackage{nicefrac}       
\usepackage{microtype}      
\usepackage{xcolor}         
\usepackage{adjustbox}

\usepackage{amsmath,amsthm,amscd,amsbsy,amssymb,latexsym,url,bm}
\usepackage{subcaption}
\usepackage{helvet}
\usepackage{courier}
\usepackage{multirow}
\usepackage{xcolor}
\usepackage{url}
\usepackage{amsmath,amscd,amsbsy,amssymb,latexsym,url,bm}
\usepackage{graphicx}
\usepackage{enumitem,balance}
\usepackage{wrapfig}
\usepackage[normalem]{ulem}
\usepackage{flafter} 
\usepackage{epigraph}
\usepackage{warning} 
\usepackage{todonotes}
\usepackage{soul}
\usepackage{listings}
\usepackage{wrapfig}
\usepackage{paralist}
\usepackage{paralist}
\usepackage[T1]{fontenc}
\usepackage{lmodern}

\graphicspath{figures/}

\definecolor{codegreen}{rgb}{0,0.6,0}
\definecolor{codegray}{rgb}{0.5,0.5,0.5}
\definecolor{codepurple}{rgb}{0.58,0,0.82}
\definecolor{backcolour}{rgb}{0.95,0.95,0.92}

\lstdefinestyle{mystyle}{
    backgroundcolor=\color{backcolour},   
    commentstyle=\color{codegreen},
    keywordstyle=\color{magenta},
    numberstyle=\tiny\color{codegray},
    stringstyle=\color{codepurple},
    basicstyle=\ttfamily\footnotesize,
    breakatwhitespace=false,         
    breaklines=true               
    captionpos=b,                    
    keepspaces=true,                 
    numbers=left,                    
    numbersep=5pt,                  
    showspaces=false,                
    showstringspaces=false,
    showtabs=false,                  
    tabsize=2
}

\newcommand{\framework}{\textsl{MALib}}
\newcommand{\argmax}{\mathop{\mathrm{argmax}}\limits}

\renewenvironment{enumerate}[1]{\begin{compactenum}#1}{\end{compactenum}}

\begin{document}

\title{MALib: A Parallel Framework for Population-based Multi-agent Reinforcement Learning} 
\author{Ming Zhou$^{1}$\footnote{The first three authors are core developers. Ming contributed to the system architecture design and development; Ziyu contributed to the algorithm implementations; Hanjing contributed to the data server decoupling.}, Ziyu Wan$^{1}$, Hanjing Wang$^{1}$, Muning Wen$^{1}$, Runzhe Wu$^{1}$, \\ Ying Wen$^{1}$\footnote{Correspondence to: Ying Wen <ying.wen@sjtu.edu.cn>.}, Yaodong Yang$^{2}$, Weinan Zhang$^{1}$ and Jun Wang$^{2}$ \\ ~ \\
$^{1}$Shanghai Jiao Tong University,
$^{2}$University College London}
\maketitle 





\begin{abstract}
Population-based multi-agent reinforcement learning (PB-MARL) refers to the series of methods nested with reinforcement learning (RL) algorithms, which produces a self-generated sequence of tasks arising from the coupled population dynamics. By leveraging auto-curricula to induce a population of distinct emergent strategies, PB-MARL has achieved impressive success in tackling multi-agent tasks. 
Despite remarkable prior arts of distributed RL frameworks, PB-MARL poses new challenges for parallelizing the training frameworks due to the additional complexity of multiple nested workloads between sampling, training and evaluation involved with heterogeneous policy interactions. To solve these problems, we present \framework{}, a scalable and efficient computing framework for PB-MARL. Our framework is comprised of three key components: (1) a centralized task dispatching model, which supports the self-generated tasks and scalable training with heterogeneous policy combinations; (2) a programming architecture named \textsl{Actor-Evaluator-Learner}, which achieves high parallelism for both training and sampling, and meets the evaluation requirement of auto-curriculum learning; (3) a higher-level abstraction of MARL training paradigms, which enables efficient code reuse and flexible deployments on different distributed computing paradigms. Experiments on a series of complex tasks such as multi-agent Atari Games show that \framework{} achieves throughput higher than $40$K FPS on a single machine with $32$ CPU cores; \textbf{$5 \times$} speedup than RLlib and at least $3\times$ speedup than OpenSpiel in multi-agent training tasks. MALib is publicly available at \url{https://github.com/sjtu-marl/malib}.
\end{abstract}

\section{Introduction}

Training intelligent 
agents that can
adapt to a diverse set of complex environments and agents has been a long-standing challenge.
A feasible way to handle these tasks is multi-agent reinforcement learning (MARL) \cite{bucsoniu2010multi}, which has shown great potentials to solve multi-agent tasks such as real-time strategy games \cite{starcraft2}, traffic light control~\cite{wu2017emergent} and ride-hailing~\cite{zhou2019multi}.
In particular, the PB-MARL algorithms combine deep reinforcement learning (DRL) and dynamical population selection methodologies 
(e.g., game theory~\cite{heinrich2016deep}, evolutionary strategies~\cite{salimans2017evolution}) to generate auto-curricula. 
In such a way, PB-MARL continually generates 
advanced intelligence and has achieved impressive successes in some non-trivial tasks, like Dota2~\cite{dota5}, StrarCraftII~\cite{vinyals2019grandmaster} and Leduc Poker~\cite{mcaleer2020pipeline}.
However, due to the intrinsic dynamics arising from multi-agent and population, these algorithms have intricately nested structure and are extremely data-thirsty, requiring a flexible and scalable training framework to ground their effectiveness.

\begin{figure}[ht!]
  \centering
  \includegraphics[width=.8\textwidth]{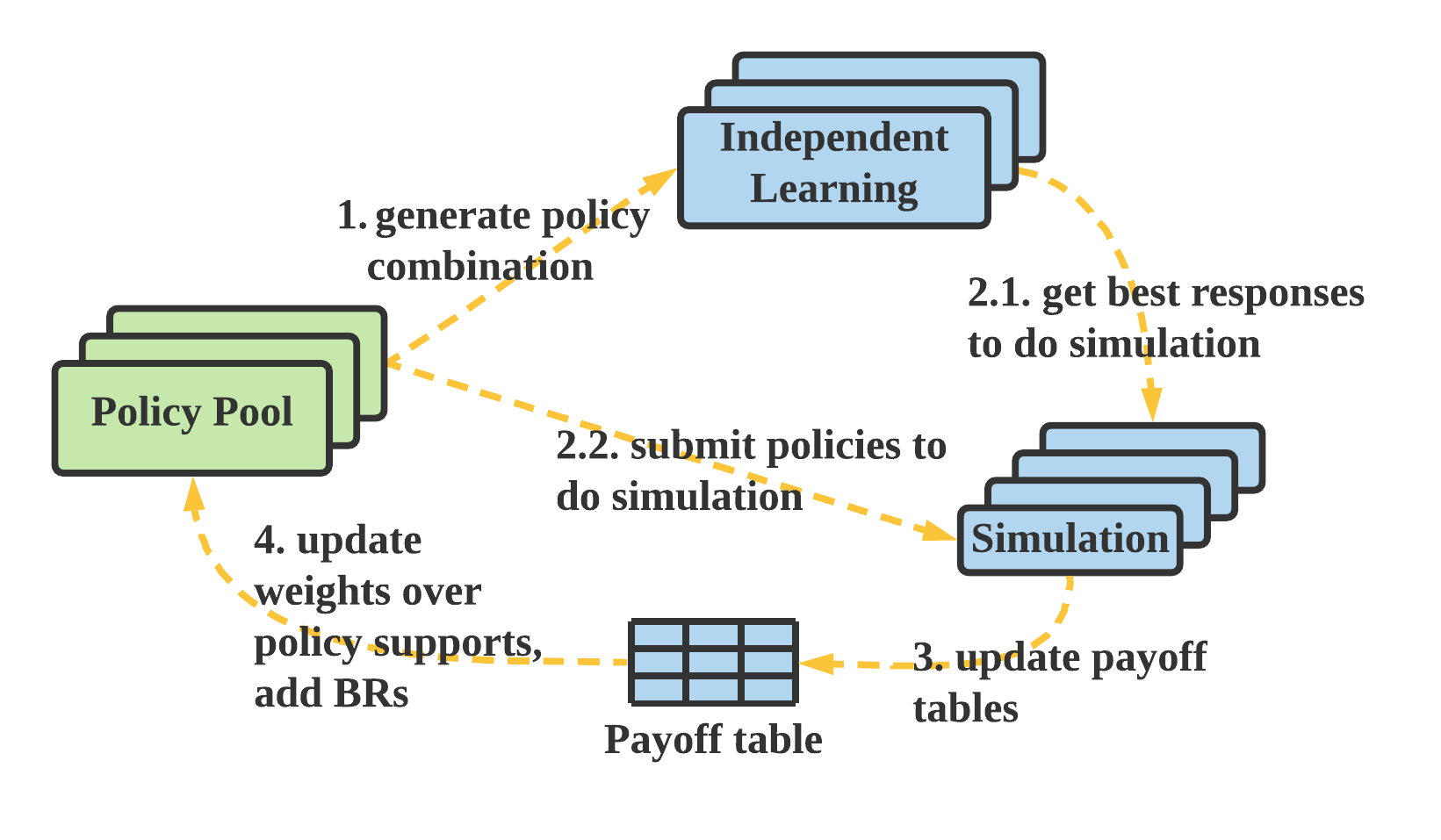}
  \captionof{figure}{PSRO learning process.}
  \label{fig:psro_learning}
\end{figure}

The deployment of PB-MARL shares some common procedures with conventional (MA)RL, but many challenges still remain, including the auto-curricula learning process and exponential compositional sampling.
PB-MARL inherently comprises heterogeneous tasks, including simulation, policy inference, policy training and policy support expansion. All of these tasks are coupled
to the underlying mutable policy combinations, which further complicates the PB-MARL.
Figure~\ref{fig:psro_learning} presents a typical case in PB-MARL, the learning process of Policy Space Response Oracle (PSRO) \cite{psro}.
At each iteration, the algorithm will generate some policy combinations and executes independent learning for each agent. Such a nested learning process comprises rollout, training, evaluation in sequence, and works circularly 
until the algorithm finds the estimated Nash Equilibrium. We note that these sub-tasks perform highly heterogeneous, i.e., the underlying policy combination is different from agent to agent. Furthermore, the evaluation stage involves tremendous simulations that cross fast-growing joint policy sets. Thus, we believe these requirements make distributed computing unavoidable for achieving high efficiency in executions.

Most of existing training frameworks are originally designed for single-agent RL, and few attempts have been made to miscellaneous types of training schema in MARL, especially PB-MARL.
In single-agent DRL, to meet the requirements of distributed computing, there have been tremendous distributed frameworks proposed to solve the data processing problem, e.g., RLlib~\cite{rllib}, SEED RL~\cite{seedrl}, Sample-Factory~\cite{petrenko2020sample} and Acme~\cite{hoffman2020acme}. 
Despite that single-agent RL methods can also be applied in multi-agent settings~\cite{dota5} by acting as independent learners~\cite{il}, more effective training schemas like centralized training \& decentralized execution~\cite{maddpg,qmix}, self-play~\cite{fsp,gwfp,psro}, population-based learning, etc. require the training performs in a non-independent style. 
Obviously, it requires extensive effort for the single-agent learning mode to handle such interactive requirements, and poses new challenges to the existing distributed reinforcement learning frameworks.
Though works like OpenSpiel~\cite{lanctot2019openspiel} have attempted to support self-play and PSRO, they merely consider the good PB-MARL abstractions and scalability. Therefore, we claim that existing frameworks cannot fully satisfy the new requirements brought by PB-MARL.

In this paper, we stress the necessity and unfulfilled requirements for deploying PB-MARL and provide our systematic solution. We summarize that the essentials of a MARL distributed framework should (1) provide a highly efficient controlling mechanism to support self-supervised auto-curricula training and heterogeneous policy interactions in PB-MARL; (2) provide a high-level abstraction of MARL training schema to simplify the implementation and a unifed and scalable training criteria. 
Based on the above analysis and comparisons, we present \framework{} as shown in Figure~\ref{fig:architecture}, a parallel framework that meets these requirements, to solve PB-MARL tasks in a high-parallelism style. We evaluate the performance of \framework{} from two distinct perspectives, i.e., system and algorithm. System performance is illustrated by the sample efficiency and throughput with increasing number of workers, while the algorithmic-side performance comparison is conducted over the reproduced algorithms with baselines, including typical PB-MARL algorithms and conventional MARL algorithms.

The main contributions of this work are summarized as follows:
\begin{enumerate}
\vspace{0.5em}
 \item We propose a centralized task dispatching model for PB-MARL, which achieves high flexibility in training tasks over auto-curricula policy combinations.
 \vspace{0.5em}
 \item We propose an independent programming model, \textsl{Actor-Evaluator-Learner}, to improve the efficiency of data processing and asynchronous execution.
 \vspace{0.5em}
 \item To improve the code reuse and compatibility of heterogeneous MARL algorithms, we abstract the MARL training schema, which also bridges the gap between single-point and multi-point optimization.
 \vspace{0.5em}
 \item We also provide mainstream MARL algorithm implementations and verify system performance and algorithm performance on \framework{} with different system settings, including single machine and clusters.
\end{enumerate}

\section{Related Work}\label{sec:related_work}
A fundamental challenge in MARL is that the agents tend to overfit other players~\cite{lanctot2017unified}, making it hard for the algorithms to achieve robust performance. To solve this problem, interacting with heterogeneous agents or diverse policies of co-players is unavoidable. PB-MARL is a feasible approach to solve this problem, and prior works include population-based training~\cite{carroll2019utility,jaderberg2019human}, self-play~\cite{vinyals2019grandmaster,fsp} and meta-game~\cite{psro,omidshafiei2019alpha}.

Another difficulty is the data processing, the same as all DRL tasks. For deep reinforcement learning, high throughput allows algorithms to achieve a faster convergence rate and high data efficiency. 
There are many distributed reinforcement learning algorithms/frameworks proposed in recent years~\cite{nair2015massively,mnih2016asynchronous,stooke2018accelerated}. Among them, a standard implementation is to design training and rollout workers in fully distributed control, i.e., the Actor-Learner model. Also, some of them try to mitigate the communication loads between CPU and GPU~\cite{babaeizadeh2016reinforcement} to improve the single-machine performance. Despite the impressive successes in distributed RL, most of them require users to do extra parallel programming to fit their custom requirements. RLlib~\cite{rllib} solved this problem by building a DRL framework on the top of Ray~\cite{ray}, work in a logically centralized control manner. Furthermore, the solution to MARL tasks among these frameworks is to model the MARL tasks as single-agent tasks, which decreases the computing efficiency in more general MARL settings since it requires heterogeneous agent/policy interaction in the training process.

Apart from the efforts in DRL, there are also tremendous works that integrate distributed computing techniques into deep learning architectures. Frameworks like Pytorch~\cite{paszke2019pytorch} and Horovod~\cite{sergeev2018horovod} implemented their distributed training logic over MPI~\cite{walker1996mpi}. General distributed tools like Ray~\cite{ray} and Fiber~\cite{zhi2020fiber} provide a universal API for building distributed applications, which relieve users from parallel programming such as MPI and OpenMP~\cite{chapman2007using}. 
There are also some works that focus on MARL implementation and abstraction~\cite{samvelyan2019starcraft,tian2017elf}, but most of them are too narrow to fit general MARL tasks, focusing on a specific domain~\cite{samvelyan19smac}. As we claimed in the aforementioned content, the distributed computing support for MARL is necessary to the wider access of this exciting area. 
\begin{table}[h]
\centering
\renewcommand{\arraystretch}{1.}
\captionsetup{type=table}
\caption{Comparison between \framework{} and existing distributed reinforcement learning frameworks from three dimensions.}
\label{table:framework_comaprison}
\begin{tabular}{cccc}
\toprule
\textbf{Framework} & \textbf{Single-Agent} & \textbf{Multi-Agent} & \textbf{Population Management} \\
\midrule
RLlib~\cite{rllib} & $\checkmark$ & $\checkmark$ & $\times$ \\
SeedRL~\cite{seedrl} & $\checkmark$ & $\times$ & $\times$ \\
Sample-Factory~\cite{petrenko2020sample} & $\checkmark$ & $\checkmark$ & $\times$ \\
\framework{} & $\checkmark$ & $\checkmark$ & $\checkmark$ \\
\bottomrule         
\end{tabular}
\end{table}

To meet the distributed computing requirements of PB-MARL, we built our \framework{} on top of Ray and provided an efficient training framework. Table~\ref{table:framework_comaprison} presents the comparison between \framework{} and exiting distributed reinforcement learning framework from three dimensions, i.e., single-agent RL support, multi-agent RL support and population management. Despite some of them support multi-agent RL algorithms, they are essentially independent learning, or require users' extra efforts to implement algorithms in other training paradigms. Furthermore, a key dimension for PB-MARL is the population management, i.e., maintaining a policy pool for each agent, support policy expansion, update policy distribution in auto-curriculum learning, etc. \framework{} considered these requirements and gives corresponding implementations.

\section{Parallel Programming Abstractions for PB-MARL}
\begin{figure}[h]
  \centering
  \includegraphics[width=\textwidth]{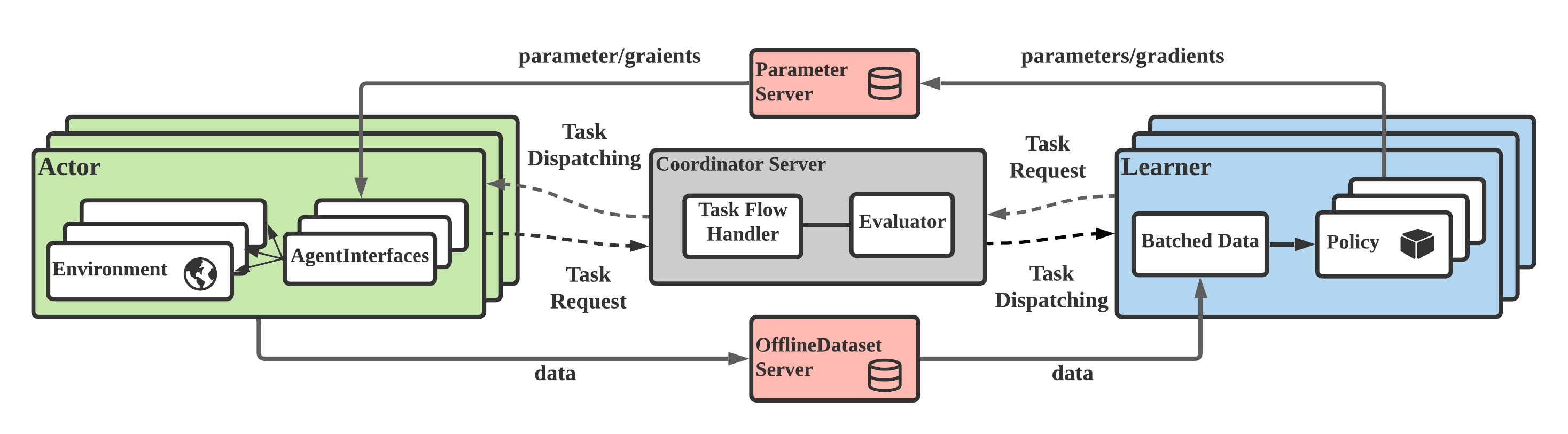}
  \caption{Overview of the \textsl{\framework{}} architecture. The \textsl{Coordinator Server} schedules the learning tasks; workers like \textsl{Actor} and \textsl{Learne}r work in parallel and data dependencies are decoupled by \textsl{Parameter} and \textsl{OfflineDataset} servers. Actor is responsible for rollout/simulation tasks with $k$ environments each, and Leaner is responsible for the optimization of a policy pool. The collected experiences are processed before being sent to the \textsl{OfflineDataset} server. After Learner/Actor completes tasks, it will send a task request to \textsl{Coordinator} server for evaluation or promote the generation of next learning stage.}
  \label{fig:architecture}
\end{figure}

In this section, we will give an introduction to our framework from three components: the \textsl{Centralized Task Dispatching} model, the \textsl{Actor-Evaluator-Learner} model and the abstractions of MARL training paradigms, as shown in Figure~\ref{fig:architecture}. With these key implementations, we tense \framework{} serve for PB-MARL in auto-curriculum learning task schedule, execution in high performance and implementation with high code reuse.

\subsection{Centralized Task Dispatching Model}
As introduced in Section~\ref{sec:related_work}, parallelizations for RL in previous work can be roughly classified into the Fully Distributed Control (FDC) ~\cite{impala,seedrl,petrenko2020sample} and the Hierarchical Parallel Task (HPT) model~\cite{rllib} fixed training task flow and policy interaction manners. Though these frameworks have abstractions for RL tasks, the extraordinary types of MARL training schema limit their performance, so users have to make extra efforts for customization.
Furthermore, the PB-MARL algorithms like PSRO~\cite{psro} and AlpahRank~\cite{omidshafiei2019alpha} require mutation in policy combination and policy space expansion in auto-curricula, which are ignored in previous frameworks.

\begin{figure}[h]
  \centering
  \includegraphics[width=\textwidth]{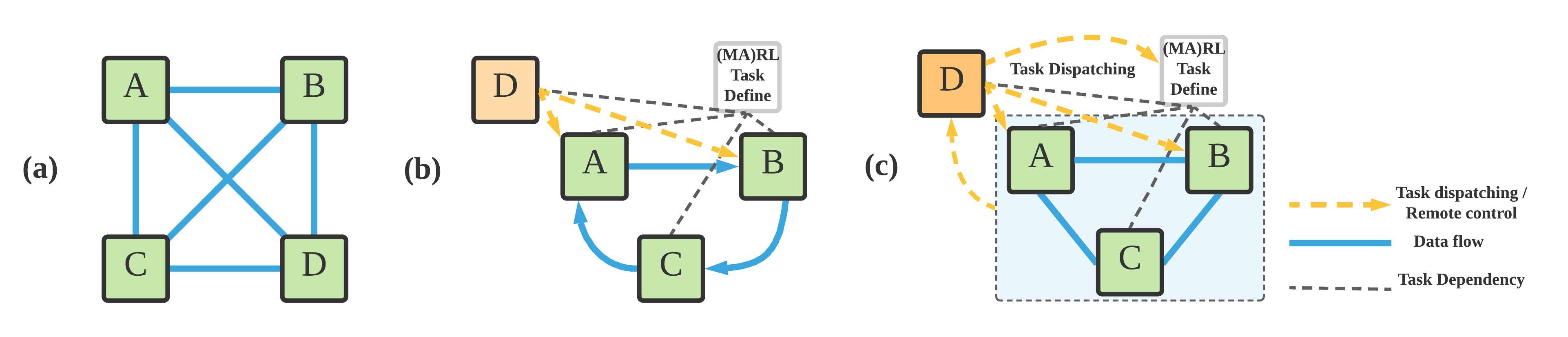}
  \caption{We abstract parallel processes as \textbf{A}$\sim$\textbf{D} in this figure. Processes perform autonomous control in (a) \textsl{Fully Distributed Control Model (FDC)} and centralized control in (b) the \textsl{Hierarchical Parallel Task Model (HPT)} and (c) our \textsl{Centralized Task Dispatching Model (CTD)}. \textbf{D} represents the centralized control process which is responsible for task dispatching. The working processes (\textbf{A}$\sim$\textbf{C}) in CTD execute in semi-passive manner, performing higher parallelism than the fully-passive execution in HPT.}
  \label{fig:control_model_comparison}
\end{figure}
We propose the \textsl{Centralized Task Dispatching (CTD)} model to meet these requirements in PB-MARL. Figure~\ref{fig:control_model_comparison} presents the comparisons between previous parallel control model and our CTD model.
This figure borrows the flowcharts from RLlib to better explain our design in parallel task control. The same as RLlib, we implemented the CTD model on top of Ray~\cite{ray}, which allows Python-implemented tasks to be naturally distributed over a large cluster.

The CTD model considers both of the advantages from FDC and HPT models. Specifically, the CTD has a centralized controller \textbf{D} to update the description of underlying policy combinations iteratively and generate new learning tasks, then dispatches them to working processes (\textbf{A}, \textbf{B} and \textbf{C}).
The working processes in CTD work independently but do not coordinate with each other like in FTD. Furthermore, the working processes also work in a semi-passive manner, i.e., they will send task requests to \textbf{D} after completing tasks, which differs from the HPT model where the working process is fully passive. In fact, the semi-passive execution can be highly performant since the working processes will not handle the centralized controller all the time, so that \textbf{D} can work in highly parallelism to process more tasks to make sure the system run in high efficiency, especially for Python, which has a global interpreter lock.
In our implementation, we modeled \textbf{D} as the \textsl{Coordinator Server}, and working processes \textbf{A}, \textbf{B} and \textbf{C} could be Actors, Learners and decoupled data servers.

\begin{figure}[ht!]
  \centering
  \includegraphics[width=\textwidth]{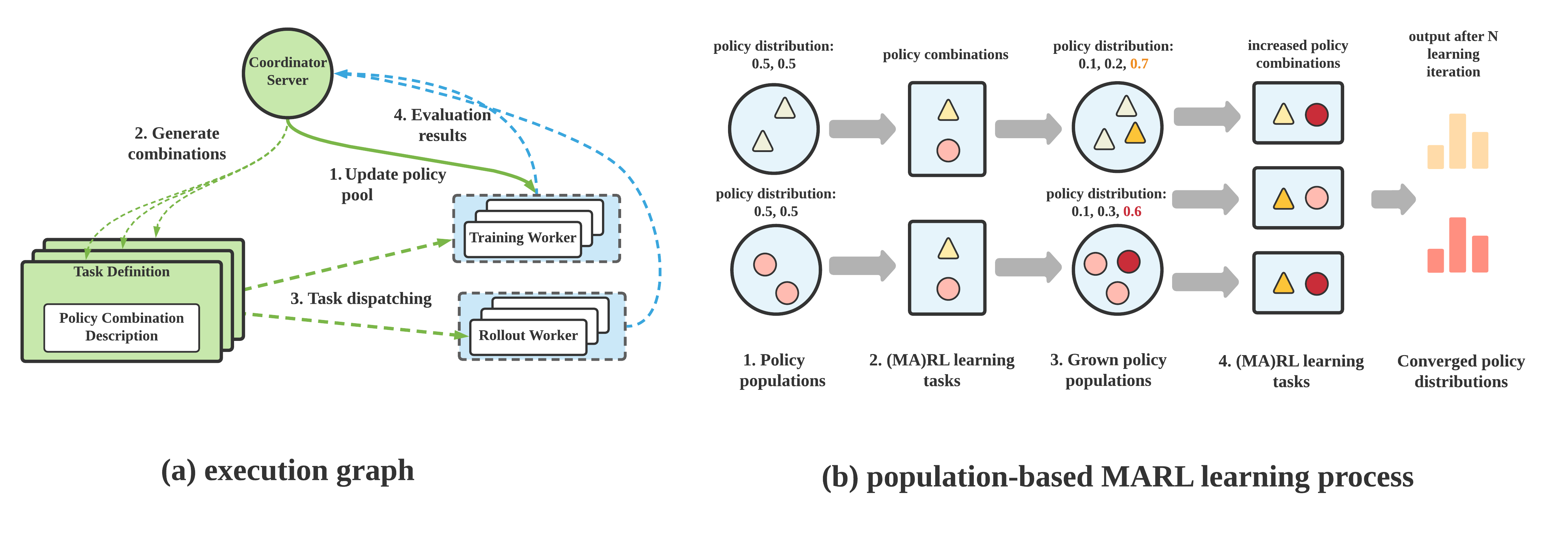}
  \caption{(a) The task execution graph of PB-MARL in \framework{}; (b) The auto-curriculum learning paradigm of PB-MARL. In general, a PB-MARL algorithm will start from an initialized policy set and the distribution over the policies (\textbf{step 1}), then find new policies (\textbf{step 2}) via (MA)RL algorithms. After that, the policy set will be updated with new policies, also the policy distribution (\textbf{step 3}), and follow with (MA)RL tasks over expanded policy combinations (\textbf{step 4}). This process will be cyclically executed until meeting the convergence such as an estimated Nash Equilibrium, then output the final policy distribution. 
 }
  \label{fig:coordination_mechanism}
\end{figure}

\paragraph{Defining the Task Graph.}
The execution logic of the CTD model can be formulated as a closed-loop task graph for PB-MARL. In the beginning, we initialize a set of policy pools $\mathcal{P} = \{\mathcal{P}_i \mid i=1,\dots, M \}$ for each agent, while some of them can share the same policy pool. Then, each agent $\{a_j \mid j=1,\dots,N\}$ from the environment will choose one policy $\pi_{a_j}$ from its belonging policy pool $\mathcal{P}_i$ to form policy combinations. We formulate a policy combination at intermediate training stage $t$ as $\text{Comb}_t=\Pi_{j}^N\pi_{a_j}$. Based on the generated $\text{Comb}_t$, the coordinator will dispatch rollout tasks and training tasks to working processes. In general, the amount of tasks is determined by specific algorithms, e.g., PSRO generates $N$ training tasks for $N$ agents if the nested RL algorithm performs independent learning, or $m \le N$ tasks for some centralized learning algorithms. Let $X(\theta_k)$ represent the collected data from rollout workers with policy parameters $\theta_k$. Then for each rollout iteration, we have:
\begin{equation*}
    X(\theta_k) \sim P(s,a,s' \mid a \sim \text{Comb}_t(\theta_k)),
\end{equation*}
where $\theta_k$ represents the policy parameters at iteration $k$. For a given policy combination, a batch of collected data will be stored as $\mathcal{D}=\{X(\theta_k) \mid k=1,\dots,h\}$, and two parallel evaluation tasks for rollout and training will executed periodically as 
\begin{equation*}
    \text{Eval}_{\text{rollout}}=f\left(X(\theta_k \leftarrow \theta')\right) \text{, and~} \theta' = \argmax_{\theta} \text{Eval}_{\text{train}}=f(X' \sim \mathcal{D} \mid \theta).
\end{equation*}
Until the global evaluator from the coordinator server reports staged stopping based on either one of them, then a new policy combination $\text{Comb}_{t+1}$ will be produced with $\text{Comb}_{t+1} \sim G(\text{Eval}_{\text{rollout}}, \text{Eval}_{\text{train}})$, where $G$ could be a specific evaluation function from PB-MARL algorithm like PSRO. Figure~\ref{fig:coordination_mechanism} shows the execution of circled task graph.

\paragraph{Decoupling the Task-Data Flow.}
The data flow mentioned here denotes the flow from data collecting to data sampling, also parameters push \& pull. In general, a data flow is private to a policy combination, and the corresponding working processes perform in high efficiency as one-to-many (one learner to multiple actors). Although such mode has shown many advantages in prior distributed frameworks~\cite{impala,seedrl,petrenko2020sample,acme}, it limits the parallelism in the PB-MARL case since the data dependencies could be many-to-many here, i.e., each rollout task is corresponding to policies from multiple learning processes whose learning paces differ to each one. We decouple the data flow from task execution using Parameter Server~\cite{li2013parameter} and OfflineDataset Server, i.e., \textsl{Parameter Server} for parameters synchronous between Actors and Learners, \text{OfflineDataset Server} for data saving and sampling, as shown in Figure~\ref{fig:architecture}. 

\subsection{Actor-Evaluator-Learner Model}
\begin{figure}[ht!]
  \centering
  \includegraphics[width=.65\textwidth]{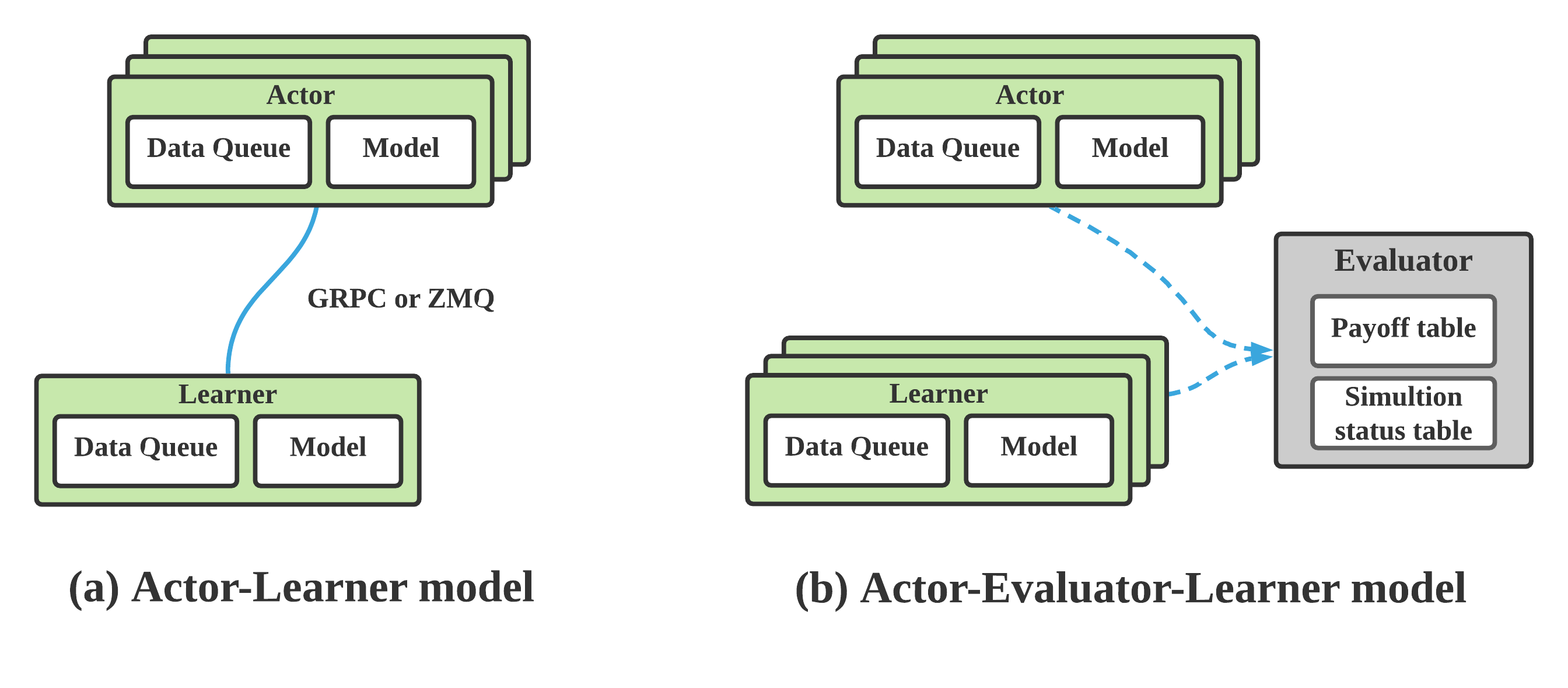}
  \caption{Comparison between Actor-Learner and Actor-Evaluator-Learner model.}
  \label{fig:al_and_ael}
\end{figure}

In single RL, it is common to decouple the training and rollout tasks~\cite{seedrl,impala,tleague,petrenko2020sample} using the \textsl{Actor-Learner} model, where the Learner operates policy training and the Actor operates data collecting. Furthermore, the evaluation program interleaves the Learners and Actors.
However, in the case of PB-MARL, such a design appears to be inadequate since there is no centralized evaluator to integrate evaluations of multiple learning tasks which is required by PB-MARL tasks. Thus, we propose the \textsl{Actor-Evaluator-Leaner} to meet this requirement, as shown in Figure~\ref{fig:al_and_ael}. The \textsl{Evaluator} is nested with a payoff table to record the evaluation results of each policy combination. In general, the evaluation of a PB-MARL algorithm is built on top of the table, e.g., PSRO generates policy distribution over existing policies by estimating a Nash Equilibrium with this table.

\subsection{Abstractions for MARL Training Paradigms}\label{sec:training_agent_interface}
PB-MARL is nested with (MA)RL algorithms. In this section, we introduce a set of learners abstracted for different MARL paradigms. Our initial implementation offers five learners to meet the basic requirements, which also considers the asynchronous and synchronous training styles and some implementations of their respective MARL algorithms.
Most previous works have tried to apply modular design to (MA)RL algorithm implementation and training. Still, they focus on the algorithm types that originate from RL, not the training paradigms, i.e., value-based or policy-gradient-based learners~\cite{acme}.
Though such a modular design presents a completed implementation logic to users, it has low reuse because of the nested training logic in algorithm implementation. We list four typical training paradigms here and present the corresponding abstractions.

\begin{figure}[t!]
    \centering
    \includegraphics[width=0.5\textwidth]{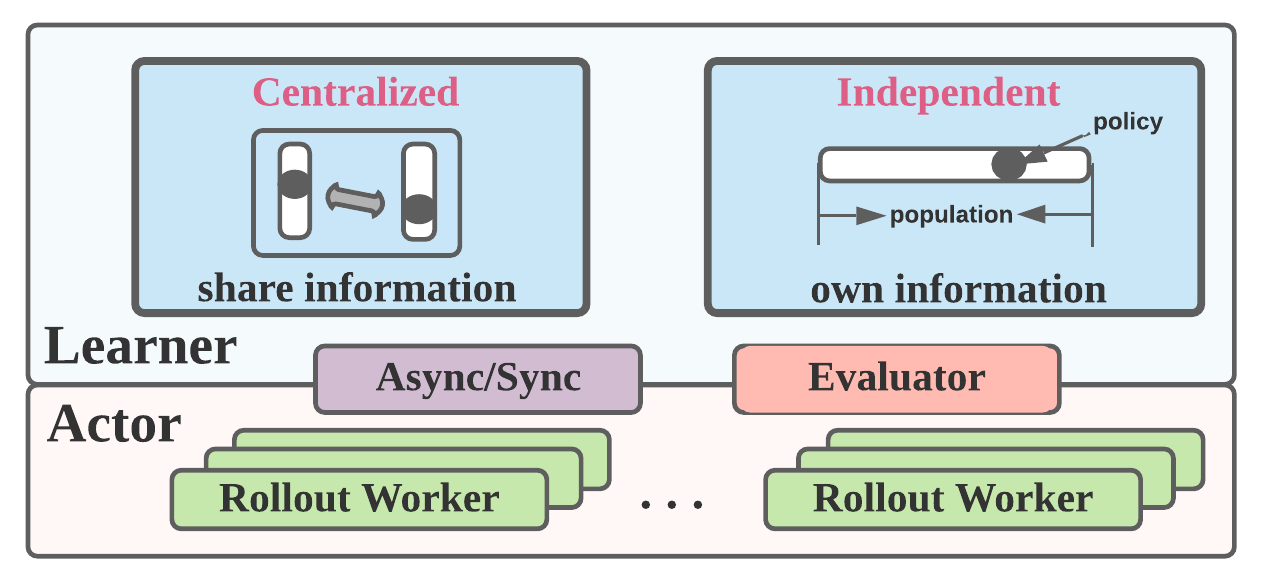}
    \caption{Depiction of scopes of independent/centralized and async/sync training diagrams. Independent and centralized paradigms have differences in data flow dependency: whether to share information among agents. And async and sync paradigms have differences in the task flow dependency: whether to strictly interleaving rollout and training.}
    \label{fig:digarams}
\end{figure}

\begin{table}[ht!]
\centering
\renewcommand{\arraystretch}{1.2}
\captionsetup{type=table}
\caption{\framework{}'s training abstraction supports various of (MA)RL and PB-MARL algorithms.}
\label{table:learner_intro}
\resizebox{0.75\linewidth}{!}{
\begin{tabular}{lcccc}
 \toprule
 \textbf{Algorithm Family} & \textbf{Independent} & \textbf{Centralized} & \textbf{Asynchronous} \\
 \midrule
 Value Based & DQN~\cite{dqn} & QMIX~\cite{qmix} & Gorrila~\cite{nair2015massively} \\
 Actor Critic & A2C~\cite{konda2000actor} & MAAC~\cite{iqbal2019actor} & IMPALA~\cite{impala} \\
 Population-based Training & PSRO~\cite{psro} & $\times$ & Pipeline-PSRO~\cite{mcaleer2020pipeline} \\
 \bottomrule
\end{tabular}
}
\end{table}

\paragraph{Independent Learning.} It is basically equivalent to single-agent deep reinforcement learning, with little or no need to change the original framework. The independent learner serves for independent learning algorithms like DQN~\cite{dqn} and PPO~\cite{ppo}. In this learning paradigm, one learner for one policy, policy learning is independent to other agents (policies).

\paragraph{Centralized Learning.} 
In MARL, centralized learning means the learning requires shared information from other agents. In MADDPG~\cite{maddpg}, the shared information indicates the state-action pairs from all agents, and each agent has its individual critic, which accepts the shared information to perform policy optimization. Another variant is QMIX~\cite{qmix}, which differs from MADDPG in that all agents share a global critic to do simultaneous policy optimization. Gradient shared algorithms like Networked-agent learning~\cite{zhang2019decentralized} could also be framed as centralized learning. Though its authors claimed that their method is fully decentralized, it is explained from the view of optimization.

\paragraph{Asynchronous Learning.}
Methods like IMPALA~\cite{impala} use multiple Learners to update policy networks in asynchronous manner. We provide an implementation to support this paradigm. Furthermore, this learning interface can be coped with the former two as a nested implementation.

\paragraph{Synchronous Learning.}
Synchronous learning interface makes the Actors and Learner work in sequence, like the conventional \textsl{Actor-Learner} model, i.e., support multiple Actors work for one Learner. Furthermore, this learning interface can be coped with the Independent and Centralized learning interfaces as a nested implementation.

We demonstrate the exiting (MA)RL algorithm families within our abstractions in Table~\ref{table:learner_intro}. The synchronous learning dimension is eliminated from the table since it is the default training manner for conventional (MA)RL algorithms. Furthermore, we show the aforementioned training paradigms in Figure~\ref{fig:digarams}.



\section{Evaluation}
\label{sec:eval}
Special efforts are paid to optimize the learning performance of multi-agent tasks in the design of \framework{}.
In this section, metrics including data throughput, sampling efficiency and training time are reported for a comprehensive understanding of \framework{} system performance. All the experiment results listed are obtained with one of the following hardware settings: (1) \textit{System\# 1}: A 32-core computing node with dual graphics cards. (2) \textit{System\# 2}: A two-node cluster with each node owns 128-core and a single graphics card. All the GPUs mentioned are of the same model (NVIDIA RTX3090). We present the primary results here, and readers can find more details in Appendix~\ref{app:results}.

\subsection{Throughput Comparisons}
We conduct the throughput evaluation on \framework{} and compare the results with some of the existing SOTA distributed RL frameworks, i.e. RLlib~\cite{rllib}, Sample-Factory~\cite{petrenko2020sample} and SEED RL~\cite{seedrl}. Specially, Sample-Factory and SEED RL are highly tailored for TPU and GPU instances. Therefore, we repeat the group of experiments, with only CPU instances and GPU acceleration activated correspondingly.

\paragraph{Environment.} As the environment for throughput comparison, we adopt the multi-agent version of Atari games (MA-Atari) from PettingZoo~\cite{terry2020pettingzoo}, which is a collection of 2D video games with more than one agent in each game. In our experiments, we use the two-players Pong game, with RGB-image frames in $12\times 12 \times 3$ resolution.
\paragraph{Baselines and Settings.} We choose IMPALA for SEED RL and RLlib, APPO for Sample-Factor and \framework{}. The evaluation of training throughput was conducted in different worker configurations over three minutes of continuous training, considering performance fluctuations caused by environment reset, data concatenation and other factors like threading lock. For each worker, we fixed the number of environments as 100. The number of workers ranges from 1 to 128 to compare the upper bound and bottleneck in parallelism performance of different frameworks.

\begin{figure}[t!]
\vspace{-3mm}
    \centering
    \includegraphics[width=\textwidth]{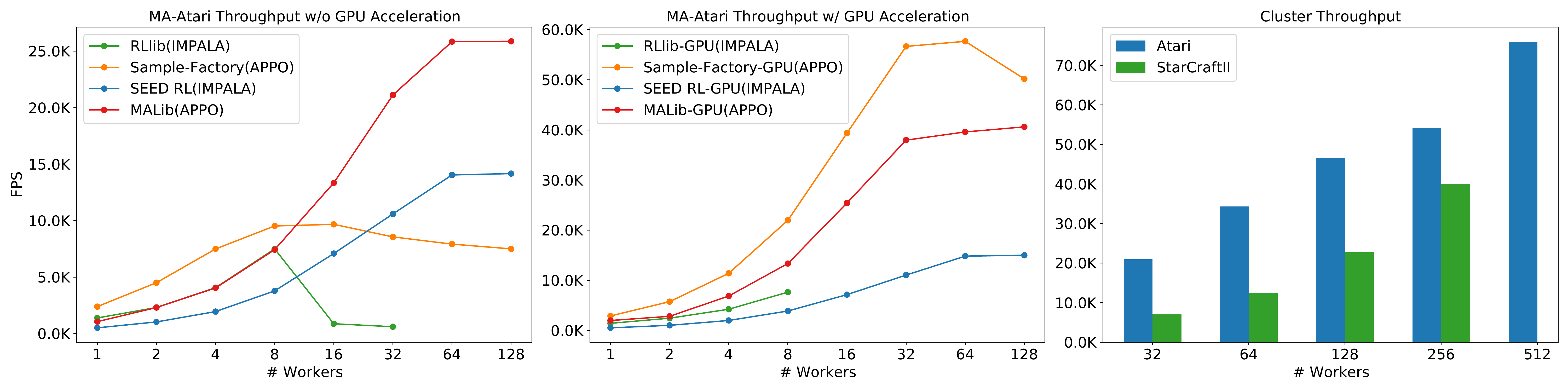}
    \caption{Throughput comparison among the existing RL frameworks and MALib. Due to resource limitation($32$ cores, $256$G RAM), RLlib fails under heavy loads( CPU case: $\#$workers$>$32, GPU case: $>$8). MALib outperforms other frameworks with only CPU and achieves comparable performance with the highly tailored framework Sample-Factory with GPU despite higher abstraction introduced. To better illustrate the scalability of \framework{}, we show the MA-Atari and SC2 throughput on system$\#2$ under different worker settings, the 512-workers group on SC2 fails due to resource limitation.
    }
    \label{fig:thp-frameworks}
\end{figure}

Figure~\ref{fig:thp-frameworks} shows the results of comparison on System\# 1. Note that due to resource limitation, with only CPU cores, RLlib failed to launch with more than $32$ workers while that threshold for GPU-accelerated RLlib is $8$ workers on the same node. Despite of the extra abstraction layer introduced for tackling PB-MARL problems, the CPU version of \framework{} outperforms other frameworks in the conventional MA-Atari environment for the scalability to different size of worker groups. And \framework{} achieves comparable performance with Sample-Factory in the GPU-acceleratd settings, which is a framework specially tailored for training conventional RL algorithms on single GPU node.

\subsection{Algorithm Implementation and Performance}
We have implemented a series of algorithms in \framework{} as listed in Appendix~\ref{app:algorithm}, including independent learning algorithms like DQN, PPO and SAC, along with MARL algorithms like MADDPG and QMIX. Furthermore, all of these algorithms can be applied to Population-based Training (PBT), self-play, which have been supported by \framework{}. For the algorithmic performance evaluation, we focus on the convergence rate, which is derived from sample efficiency and training time consumption. As for the evaluated algorithms, we use PSRO training for PB-MARL, MADDPG and QMIX for conventional MARL algorithms. With the consideration of fairness, we make the algorithm implementation from different frameworks consistent in network settings. Appendix~\ref{app:results} presents the details.

\begin{figure}[ht!]
    \centering
    \begin{subfigure}[b]{0.32\textwidth}
    \centering
        \includegraphics[width=0.93\textwidth]{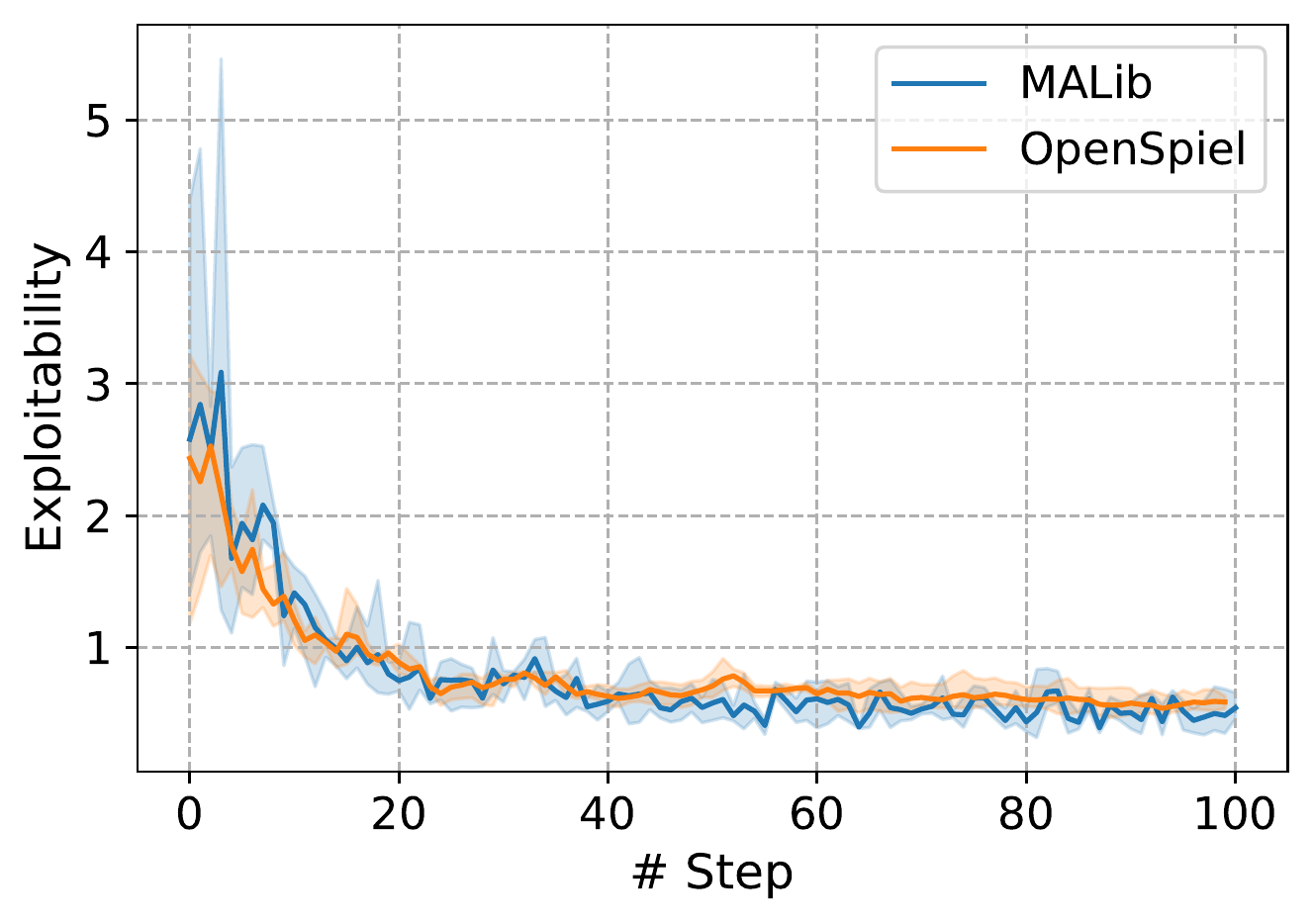}
        \vspace{-0.2cm}
        \caption{}
        \label{fig:leduc_step_time_sim2000}
    \end{subfigure}
    \hfill
    \hspace{-2.4cm}
    \begin{subfigure}[b]{0.33\textwidth}
        \centering
        \includegraphics[width=0.904\textwidth]{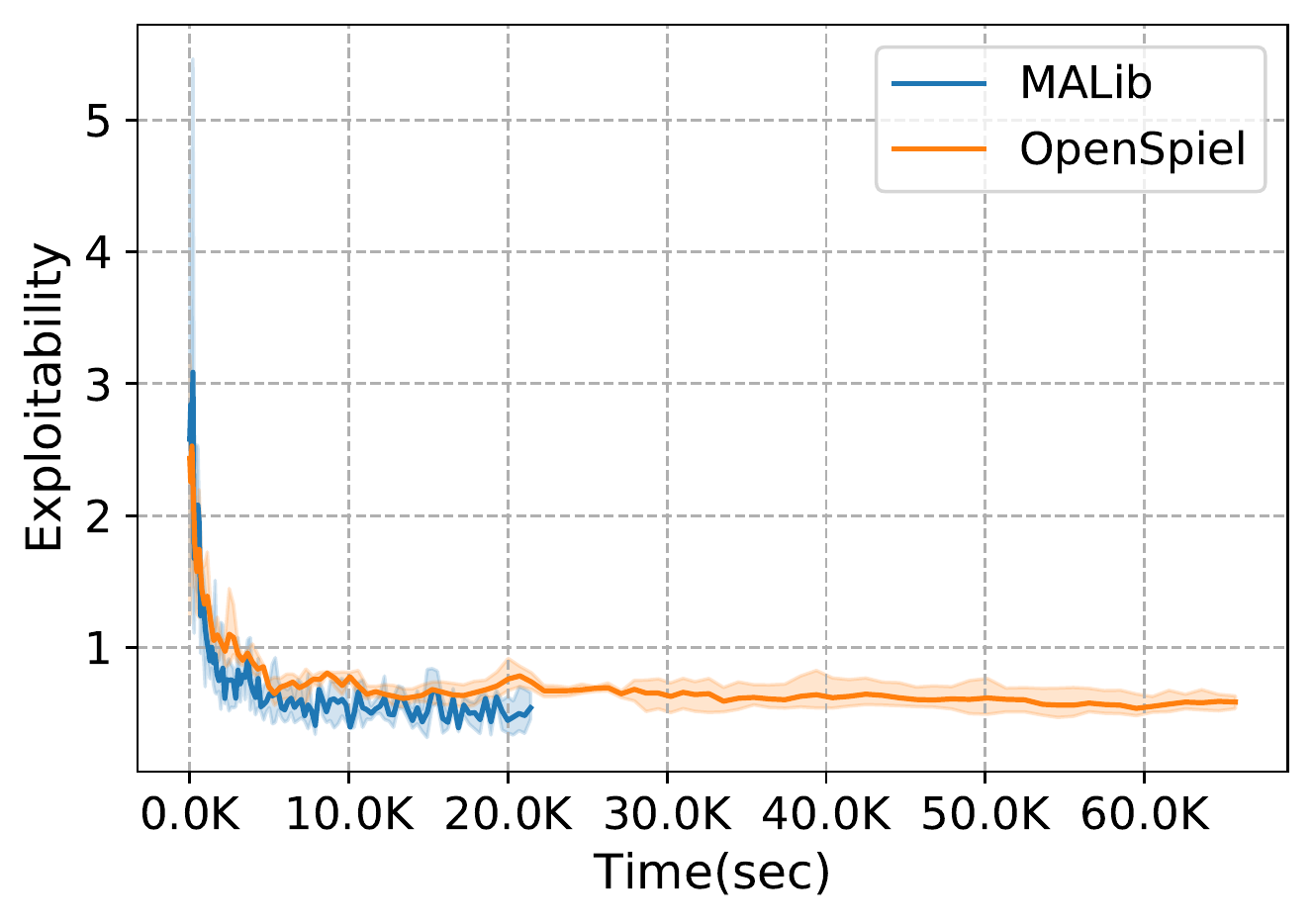}
        \vspace{-0.2cm}
        \caption{}
        \label{fig:psro_time_exp_sim2000}
    \end{subfigure}
    \hfill
    \hspace{-2.4cm}
    \begin{subfigure}[b]{0.34\textwidth}
        \centering
            \includegraphics[width=0.943\textwidth]{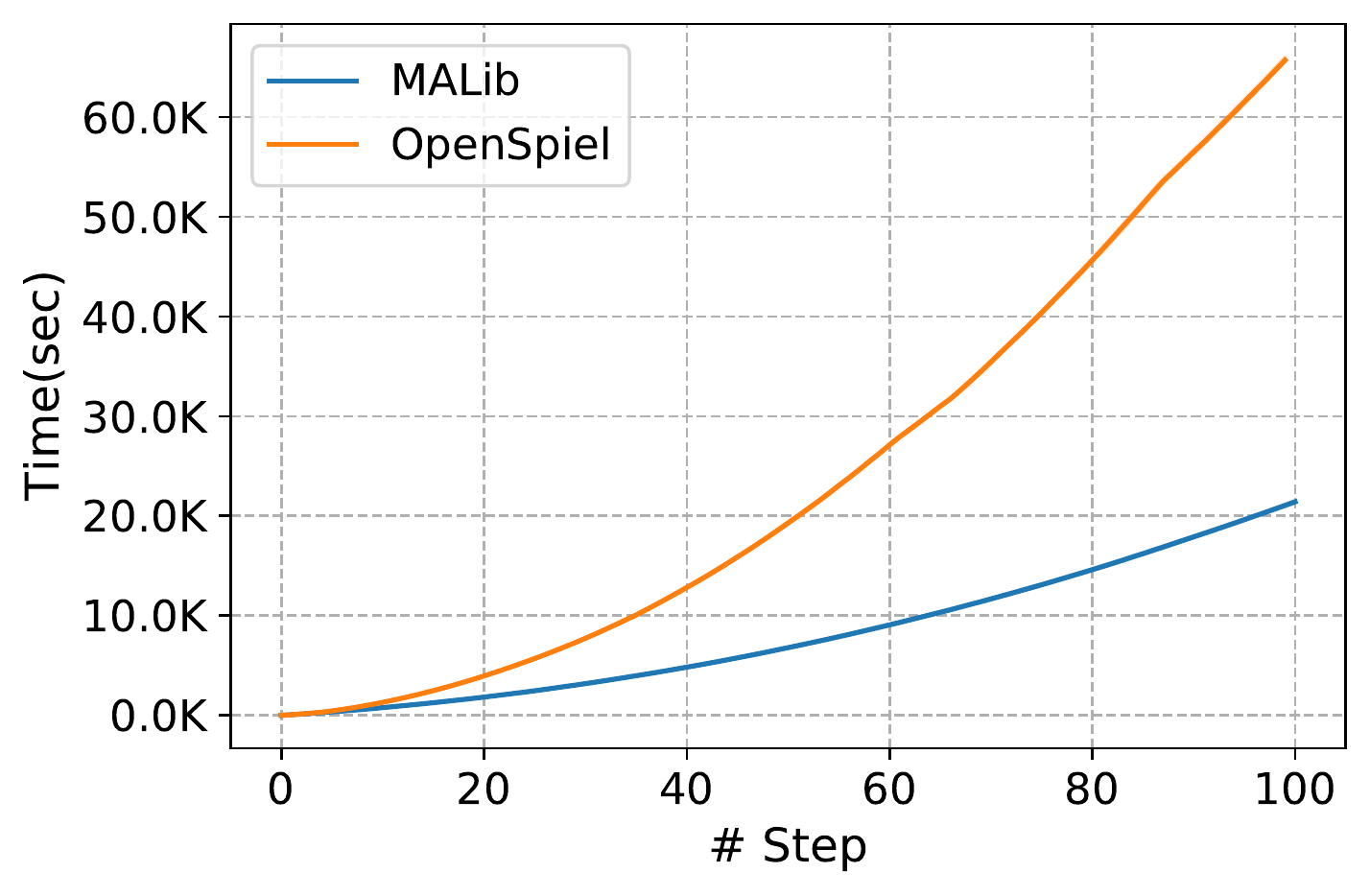}
            \vspace{-0.2cm}
            \caption{}
            \label{fig:leduc_step_time_sim2000}
    \end{subfigure}
    \vspace{-0.1cm}
    \caption{Comparisons of PSRO between \framework{} and OpenSpiel. (a) indicates that \framework{} achieves the same performance on exploitability as OpenSpiel; (b) shows that the convergence rate of \framework{} is 3$\times$ faster than OpenSpiel; (c) shows that \framework{} achieves a higher execution efficiency than OpenSpiel, since it requires less time consumption to iterate the same learning steps, which means \framework{} has the potential to scale up in more complex tasks that need run for much more steps.}
    \label{fig:psro_leduc_poker_sim2000}
\end{figure}

\begin{table}[t!]
\centering
\renewcommand{\arraystretch}{1}
\captionsetup{type=table}
\caption{\framework{}'s implementation result of population-based methods.}
\label{table:pbt_result}
\begin{tabular}{lccc}
 \toprule
 \textbf{Algorithm} & \textbf{Time(sec)} & \textbf{Population Size} \\
 \midrule
 Fictitious Self-Play & 7562& 60\\
 PSRO(fictitious play) & 4862 & 40\\
 PSRO($\alpha$-rank) & \textbf{1776} & \textbf{21}\\
 \bottomrule
\end{tabular}
\end{table}

\paragraph{Population-based training for Leduc Poker.} 
We compared the PSRO algorithm with OpenSpiel's~\cite{lanctot2019openspiel} implementation on Leduc Poker, a common benchmark in Poker AI. In order to stress how the evaluation effects the running time, we change the meta-solver to fictitious play, an approximation of Nash Equilibrium with tractable solving time on a large payoff matrix. To get a relatively accurate empirical payoff, the number of simulations was sat to be 2000 for each policy combination and the learning iteration was limited to 100 steps, i.e., the maximum of population size is 100.

We evaluate the convergence of PSRO with exploitability~\cite{mcaleer2020pipeline}. As shown in Figure~\ref{fig:psro_time_exp_sim2000}, we cut $70\%$ execution time while maintaining exploitability with a similar quality as OpenSpiel(Figure~\ref{fig:psro_leduc_poker_sim2000}).  Moreover, it indicates that \framework{} is more capable of complicated games since the executing time of OpenSpiel grows much faster than \framework{}. Other methods like Fictitious Self-Play~\cite{fsp} (FSP) and Self-Play~\cite{} (SP) were implemented and evaluated in the same environment settings. In Table~\ref{table:pbt_result}, we compare the execution time required and population size expanded when the exploitability goes to 0.5. The results show that PSRO outperforms the other two methods. Specifically, SP fails to converge since this Poker game is not a purely transitive game, and FSP is more time-consumption than PSRO. It might be that PSRO considers the interactions and meta-game between different policies in  populations, and solves it to approximate the Nash Equilibrium of the underlying game, which results in faster convergence rate. Furthermore, when an exact meta-game solver such as the LP-solver or $\alpha$-rank, PSRO will converge in shorter time and smaller population size.

\paragraph{MADDPG for MPE.} Multi-agent Particle Environments~\cite{maddpg} (MPE) is a typical benchmarking environment for MARL algorithms. It offers plenty of scenarios covering cooperative and competitive tasks. We compared MADDPG with RLlib implementation on seven scenarios under different worker settings, covered cooperative, competitive, and mixed cooperative-competitive tasks.

\begin{figure}[ht!]
    \centering
    \includegraphics[scale=0.326]{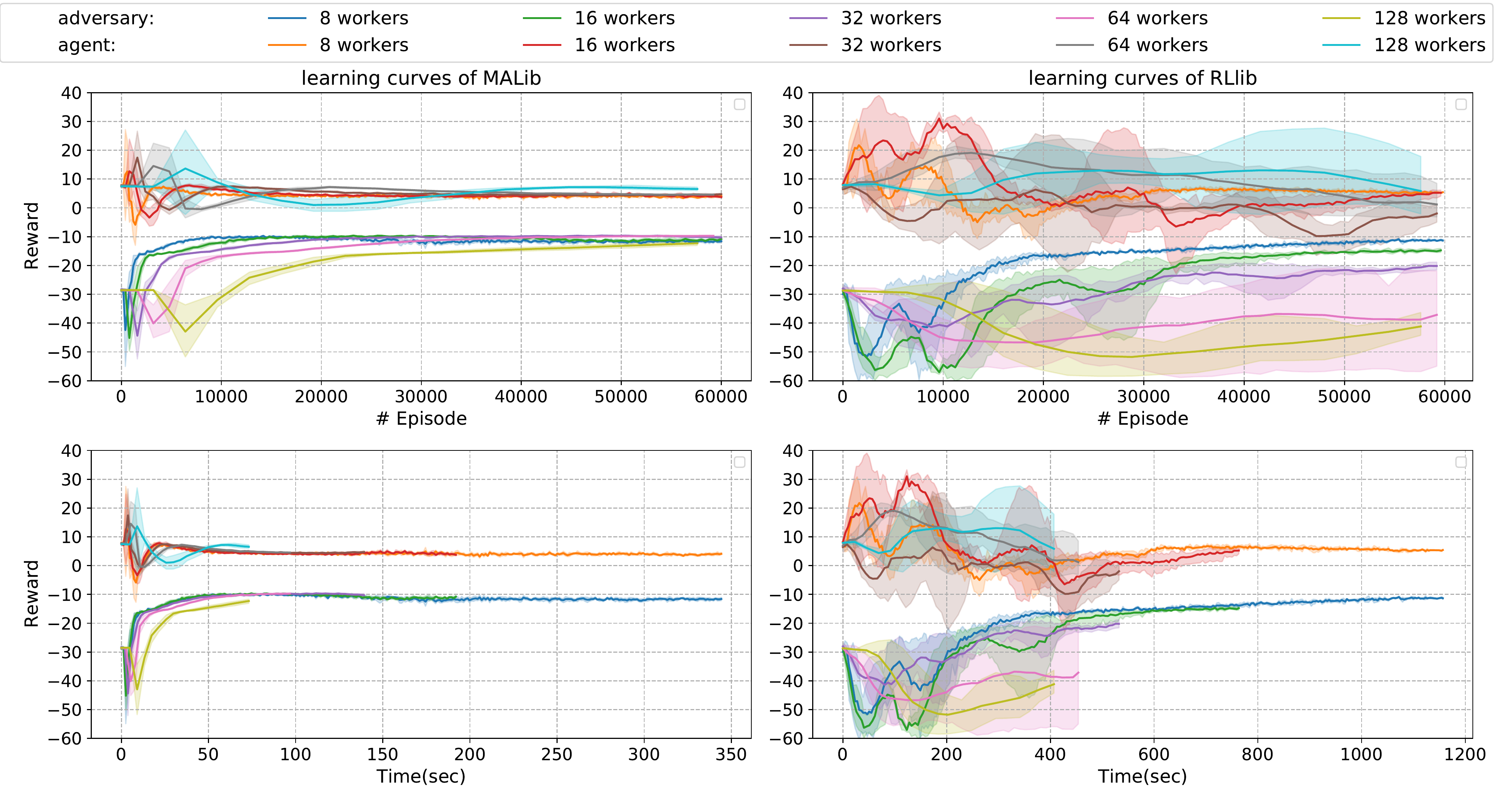}
    \caption{Comparisons of MADDPG in \textit{simple adversary} under different rollout worker settings. Figures in the top row depict each agent's episode reward w.r.t. the number of sampled episodes, which indicate that \framework{} converges faster than RLlib with equal sampled episodes. Figures in the bottom row show the average time and average episode reward at the same number of sampled episodes, which indicates that \framework{} achieves 5$\times$ speedup than RLlib.}
    \label{fig:mpe_adversary}
    \vspace{-3mm}
\end{figure}

The experiments were conducted under different worker settings to compare the convergence performance. Figure~\ref{fig:mpe_adversary} shows the results on \textit{simple adversary}, it indicates that \framework{}'s implementation performs more steadily than RLlib's, especially when the worker number increases, RLlib implementation show high variance and fail to converge. We point out that the difference in implementations between \framework{} and RLlib is that the former executes learning in a fully asynchronous and the RLlib's implementation executes learning in sequential. Despite the data sampling executes in parallel, RLlib requires extra efforts on tuning to solve the data starvation in the training stage.

\section{Conclusion}\label{sec:conclusion}
In this paper, we introduced \framework{} for PB-MARL to boost the research from high efficient training and code implementation, also compared with novel distributed RL frameworks in system and algorithm performance. Despite that \framework{} is currently built upon some of the low-level stacks from Ray \cite{ray} and lacks further optimization for GPU, it achieved higher throughput than previous work on multi-agent Atari tasks, especially in the CPU-only setting. For the algorithm performance, \framework{} shows at least 3x speedup with limited simulation parallelism on PB-MARL tasks compared to the existing library and achieved state-of-the-art scores on MPE tasks 5x faster than the parallel implementation in RLlib. System development is a long-term work, we plan to further improve MALib by optimizing the usage of the heterogeneous computing resources and testing the performance on much larger scale clusters in the future.

\section*{Acknowledgments}
The SJTU team is supported by ``New Generation of AI 2030'' Major Project (2018AAA0100900), Shanghai Municipal Science and Technology Major Project (2021SHZDZX0102), National Natural Science Foundation of China (62076161, 61632017) and Shanghai Sailing Program (21YF1421900). We also thank Zhicheng Zhang, Hangyu Wang, Ruiwen Zhou, Weizhe Chen, Minghuan Liu, Yunfeng Lin, Xihuai Wang, Derrick Goh and Linghui Meng for many helpful discussions, suggestions and comments on the project, and Ms. Yi Qu for her help on the design work.

\bibliography{main}
\bibliographystyle{plain}
\clearpage
\newpage
\appendix
\section{Algorithm Library}
\label{app:algorithm}
We have integrated a set of popular (MA)RL algorithms. Table~\ref{table:algorithms} gives an overview of these algorithms and tags them according to 1) training interface introduced in Section~\ref{sec:training_agent_interface}, 2) execution mode, and 3) the supported PB-MARL algorithms. The training interfaces could be \textbf{Independent} or \textbf{Centralized} which are corresponding to independent learning and centralized learning respectively. The execution mode could be Async (asynchronous) or Sync (synchronous). In the initial implementation, we provided three PB-MARL algorithms support, they are Policy Space Response Oracle~\cite{psro} (PSRO), Fictitious Self-play~\cite{fsp} (FSP), Self-play~\cite{heinrich2016deep} (SP) and Population-based Training~\cite{jaderberg2017population} (PBT).

\begin{table}[ht!]
	\centering
	\renewcommand{\arraystretch}{1.}
	\captionsetup{type=table}
	\caption{Implemented algorithms in \framework{}.}
	\label{table:algorithms}
	\resizebox{0.75\linewidth}{!}{
		\begin{tabular}{lccc}
			\toprule
			\textbf{Algorithm} & \textbf{Training Interface} & \textbf{Execution Mode} & \textbf{PB-MARL Support} \\
			\midrule
			DQN~\cite{dqn} & Independent & Async/Sync & PSRO/FSP/SP \\
			Gorilla~\cite{nair2015massively} & Independent & Async & PSRO/FSP/SP \\
			A2C~\cite{konda2000actor} & Independent & Sync & PSRO/FSP/SP \\
			A3C~\cite{mnih2016asynchronous} & Independent & Async & PSRO/FSP/SP \\
			SAC~\cite{haarnoja2018soft} & Independent & Async/Sync & PSRO/FSP/SP \\
			DDPG~\cite{DBLP:journals/corr/LillicrapHPHETS15} & Independent & Async/Sync & PSRO/FSP/SP \\
			PPO~\cite{ppo} & Independent & Sync & PSRO/FSP/SP \\
			APPO & Independent & Async & PSRO/FSP/SP\\
			\midrule
			MADDPG~\cite{maddpg} & Centralized & Async/Sync & PBT \\
			QMIX~\cite{qmix} & Centralized & Async/Sync & PBT \\
			MAAC~\cite{iqbal2019actor} & Centralized & Async/Sync & PBT \\
			\bottomrule
		\end{tabular}
	}
\end{table}

\section{Additional Results}
\label{app:results}




\subsection{MADDPG for MPE}
\label{app:maddpg_mpe}
We implemented MADDPG in \framework{} with the same configuration as RLlib, i.e., both of the actor and critic uses three layers of 64-units fully-connect network. The experiments were conducted in seven scenarios introduced in PettingZoo~\cite{terry2020pettingzoo}, with different worker settings (ranges from 1 to 128), as listed below.

\begin{figure}[ht!]
    \centering
    \includegraphics[scale=0.326]{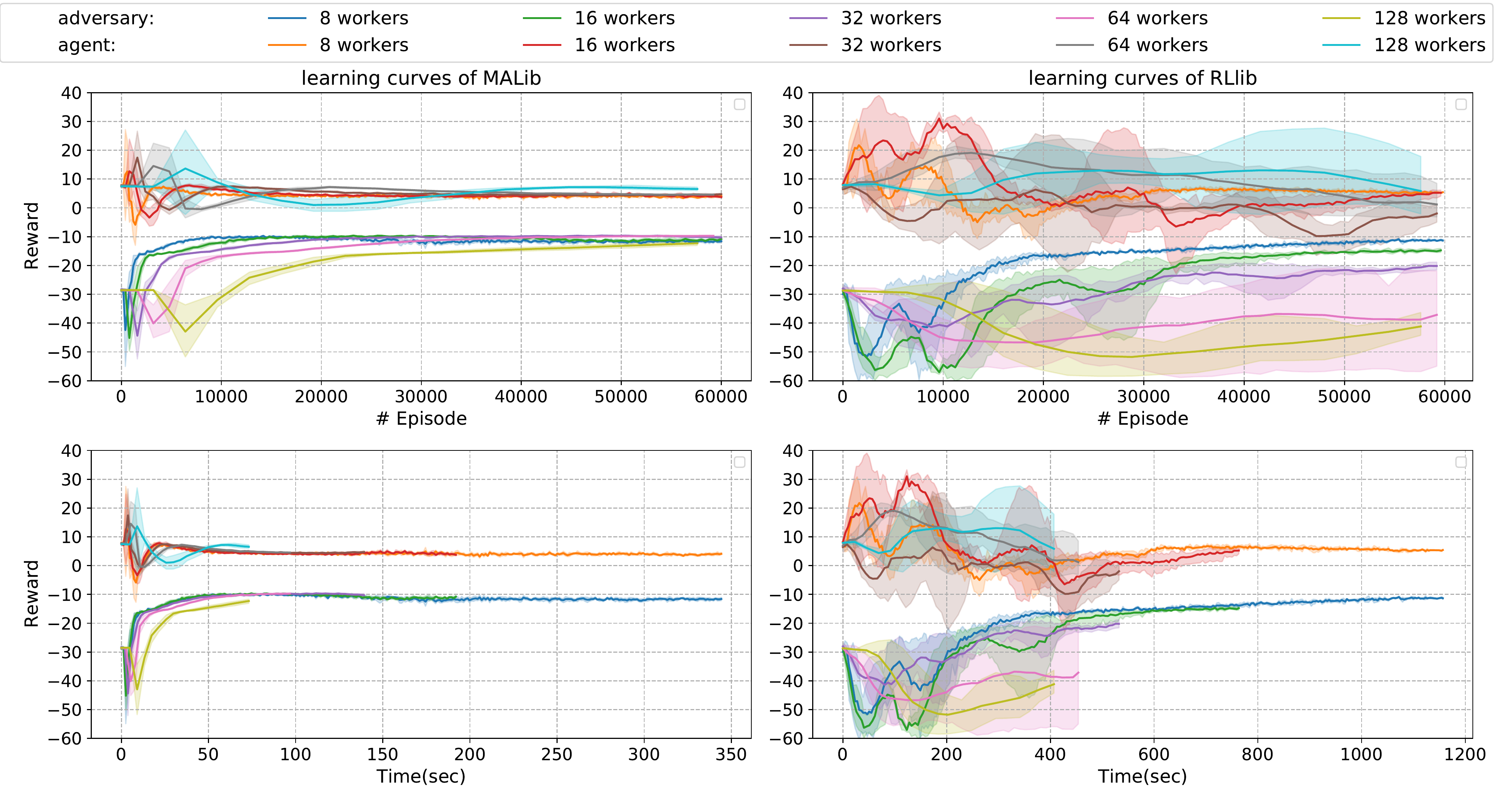}
    \caption{Comparisons of MADDPG in \textit{simple adversary} under different rollout worker settings.}
    \label{fig:mpe_adversary_2}
\end{figure}

\paragraph{Simple Adversary.} There are 1 adversary, 2 good agents and 2 landmarks in this scenario. All agents can observe landmarks and other agents. One landmark is tagged as the `target landmark'. Good agents are rewarded based on the distance to the target landmark, i.e., closer to the target landmark higher reward, but also receive negative reward based on how close the adversary is to the target landmark. For the adversary, it is rewarded based on distance to the target, but it doesn’t know which landmark is the target landmark. In this scenario, good agents have to learn to `split up' and cover all landmarks to deceive the adversary. Figure~\ref{fig:mpe_adversary_2} shows the comparison from the converged reward and time-consumption.

\begin{figure}[ht!]
    \centering
    \includegraphics[scale=0.326]{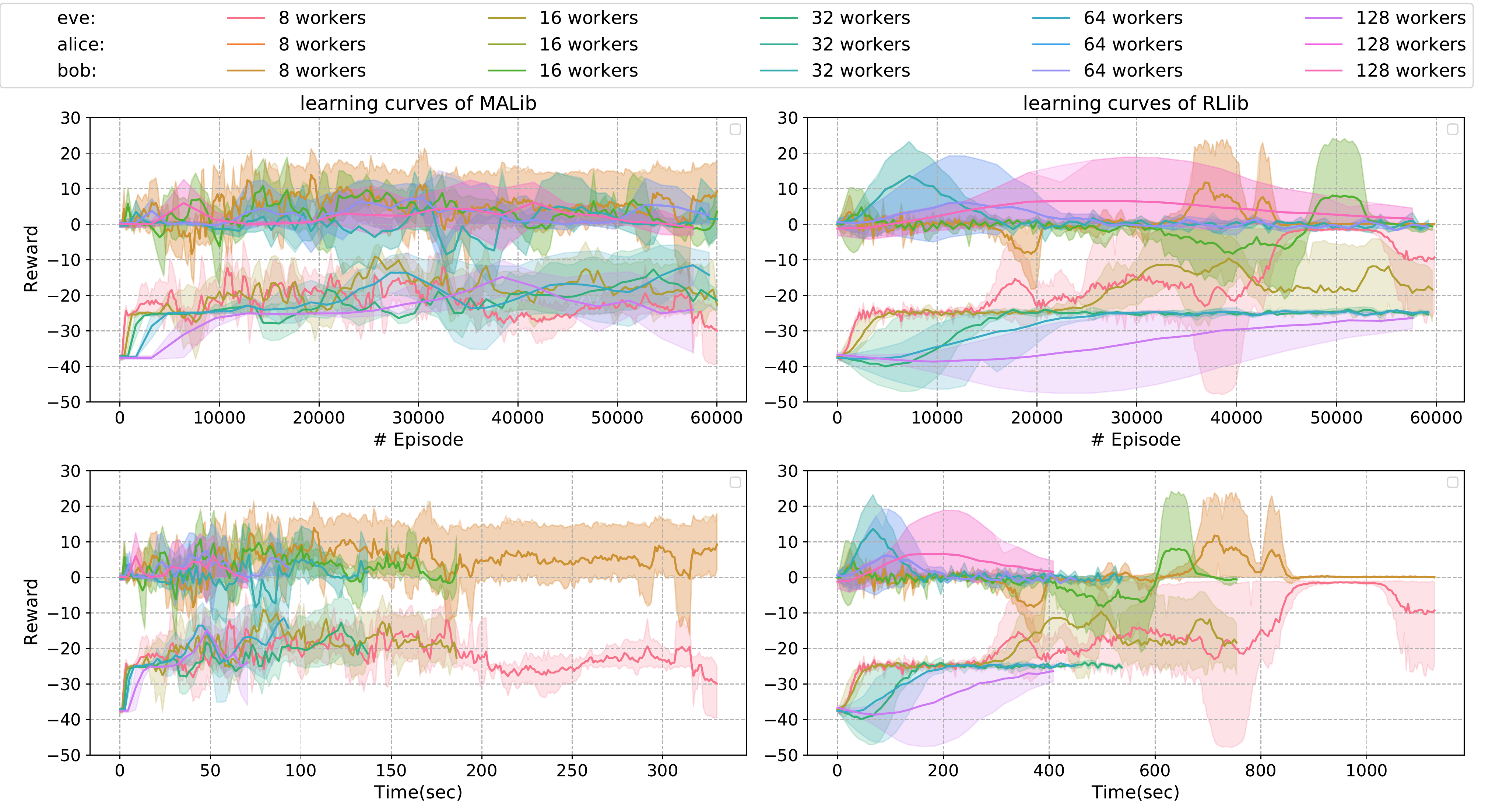}
    \caption{Comparisons between learning curves of MALib and RLlib when running MADDPG in \textit{simple crypto} environment under different rollout worker settings.}
    \label{fig:mpe_crypto}
\end{figure}

\paragraph{Simple Crypto.} There are 2 good agents (Alice and Bob) and 1 adversary (Eve) in this scenario. Alice must sent a private 1 bit message to Bob over a public channel. Alice and Bob are rewarded +2 if Bob reconstructs the message, but are rewarded -2 if Eve reconstruct the message (that adds to 0 if both teams reconstruct the bit). Eve is rewarded -2 based if it cannot reconstruct the signal, zero if it can. Alice and Bob have a private key (randomly generated at beginning of each episode) which they must learn to use to encrypt the message. Figure~\ref{fig:mpe_crypto} shows the comparison from the converged reward and time-consumption.

\begin{figure}[ht!]
    \centering
    \includegraphics[scale=0.326]{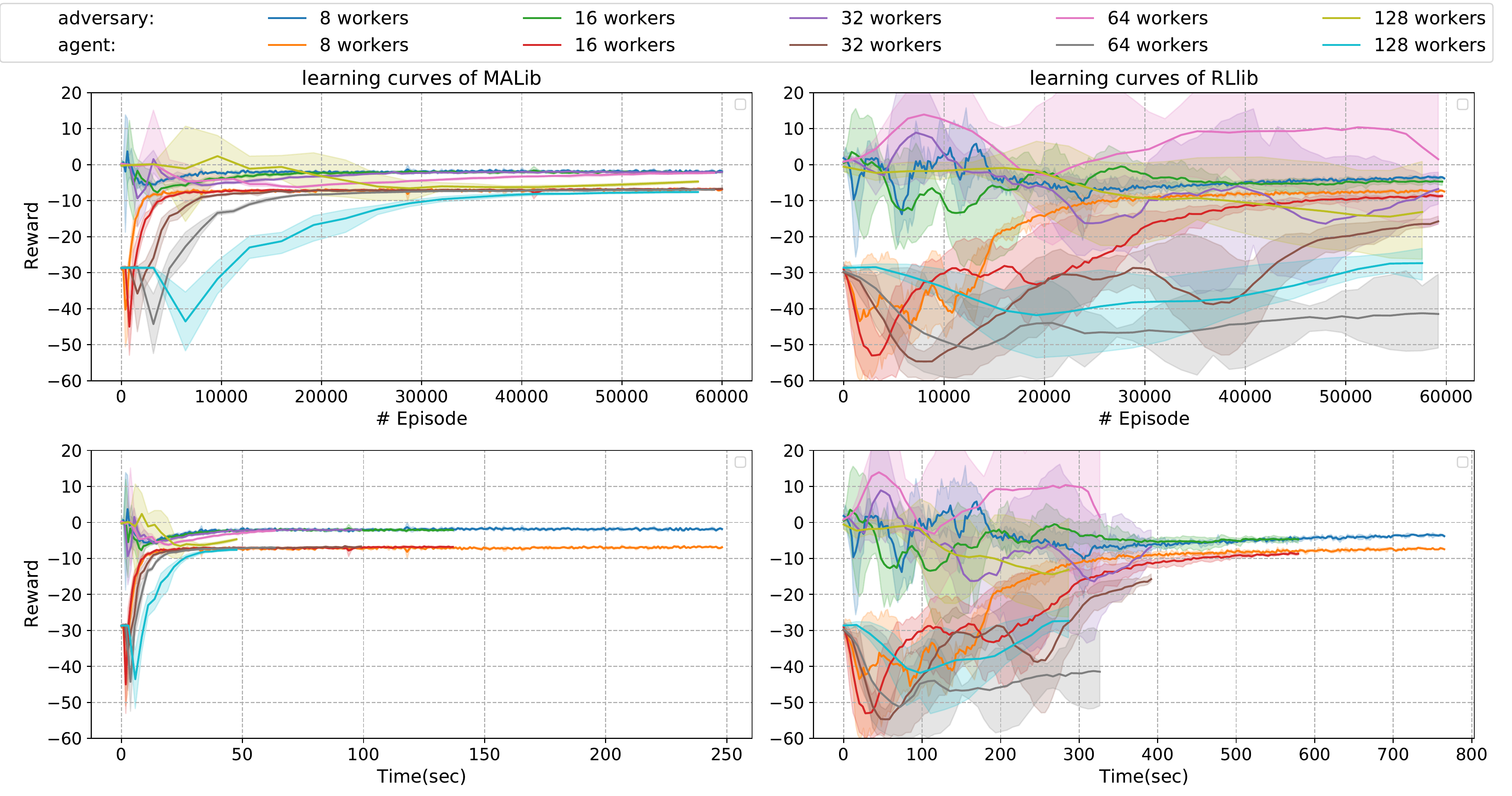}
    \caption{Comparisons between learning curves of \framework{} and RLlib when running MADDPG in \textit{simple push} under different rollout worker settings.}
    \label{fig:mpe_push}
\end{figure}

\paragraph{Simple Push.} There are 1 good agent, 1 adversary, and 1 landmark in this scenario. The good agent is rewarded based on the distance to the landmark. The adversary is rewarded if it is close to the landmark, and if the agent is far from the landmark (the difference of the distances). Thus the adversary must learn to push the good agent away from the landmark. Figure~\ref{fig:mpe_push} shows the comparison from the converged reward and time-consumption.

\begin{figure}[ht!]
    \centering
    \includegraphics[scale=0.326]{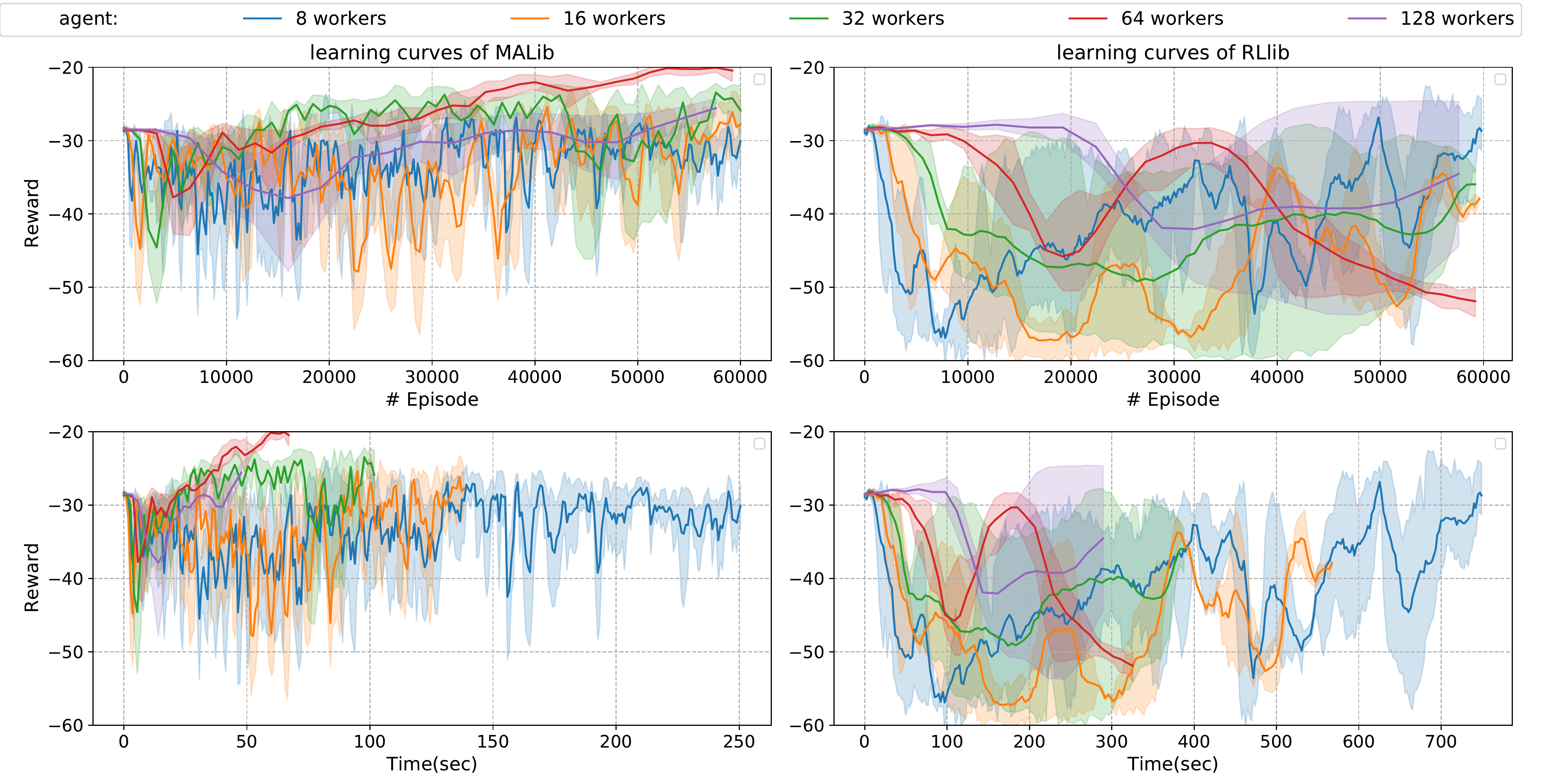}
    \caption{Comparisons between learning curves of MALib and RLlib when running MADDPG in \textit{simple reference} under different rollout worker settings.
    }
    \label{fig:mpe_reference}
\end{figure}

\paragraph{Simple Reference.} There are 2 agents and 3 landmarks of different colors in this scenario. Each agent wants to get closer to their target landmark, which is known only by the other agents. Both agents are simultaneous speakers and listeners. Locally, the agents are rewarded by their distance to their target landmark. Globally, all agents are rewarded by the average distance of all the agents to their respective landmarks. The relative weight of these rewards is controlled by the local\_ratio parameter. Figure~\ref{fig:mpe_reference} shows the comparison from the converged reward and time-consumption.

\begin{figure}[ht!]
    \centering
    \includegraphics[scale=0.326]{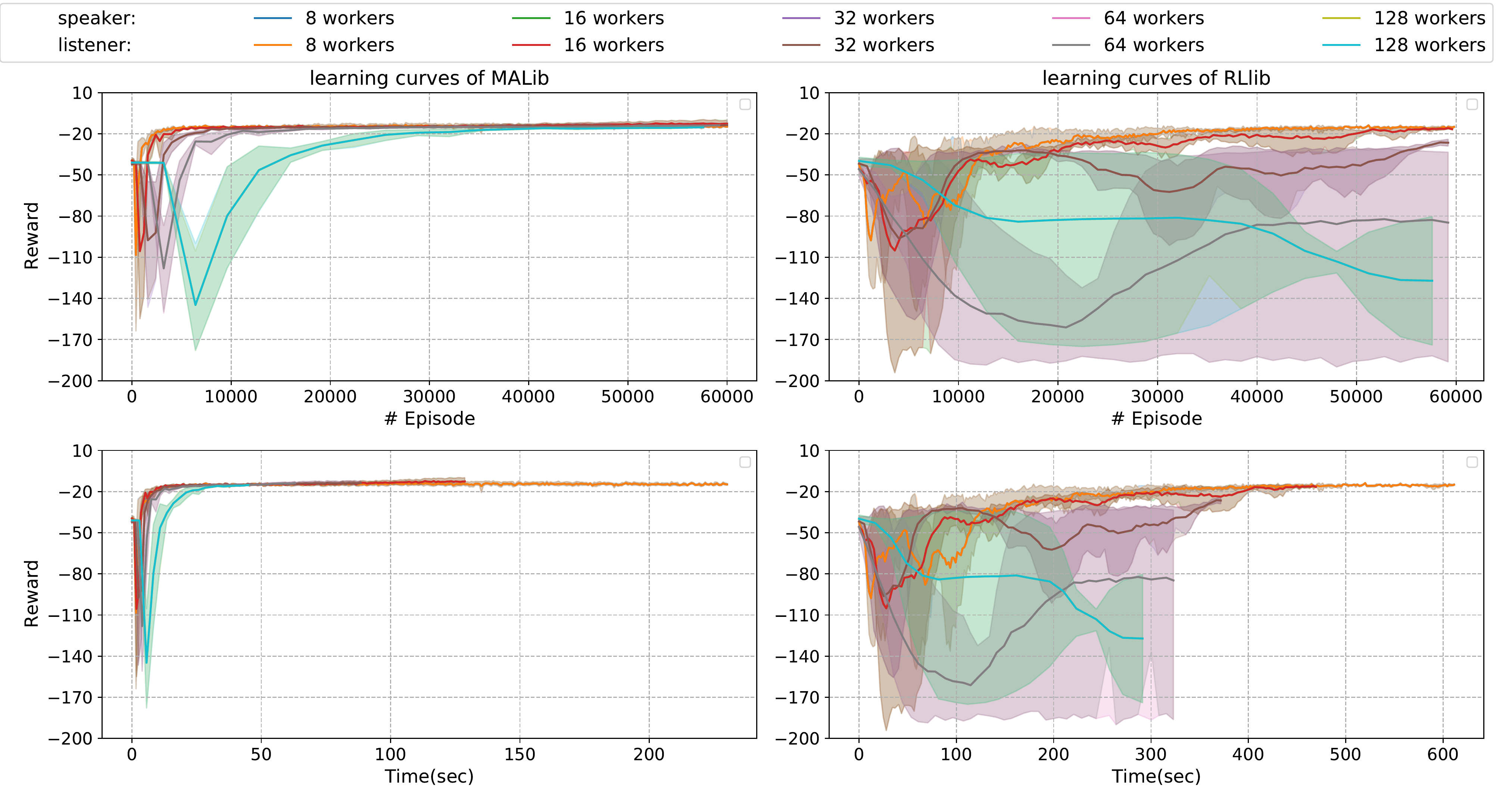}
    \caption{Comparisons between learning curves of MALib and RLlib when running MADDPG in \textit{simple speaker listener} under different rollout worker settings.
    }
    \label{fig:mpe_speaker_listener}
\end{figure}

\paragraph{Simple Speaker Listener.} This scenario is similar to simple\_reference, except that one agent is the `speaker' and can speak but cannot move, while the other agent is the listener (cannot speak, but must navigate to correct landmark). Figure~\ref{fig:mpe_speaker_listener} shows the comparison from the converged reward and time-consumption.

\begin{figure}[ht!]
    \centering
    \includegraphics[scale=0.326]{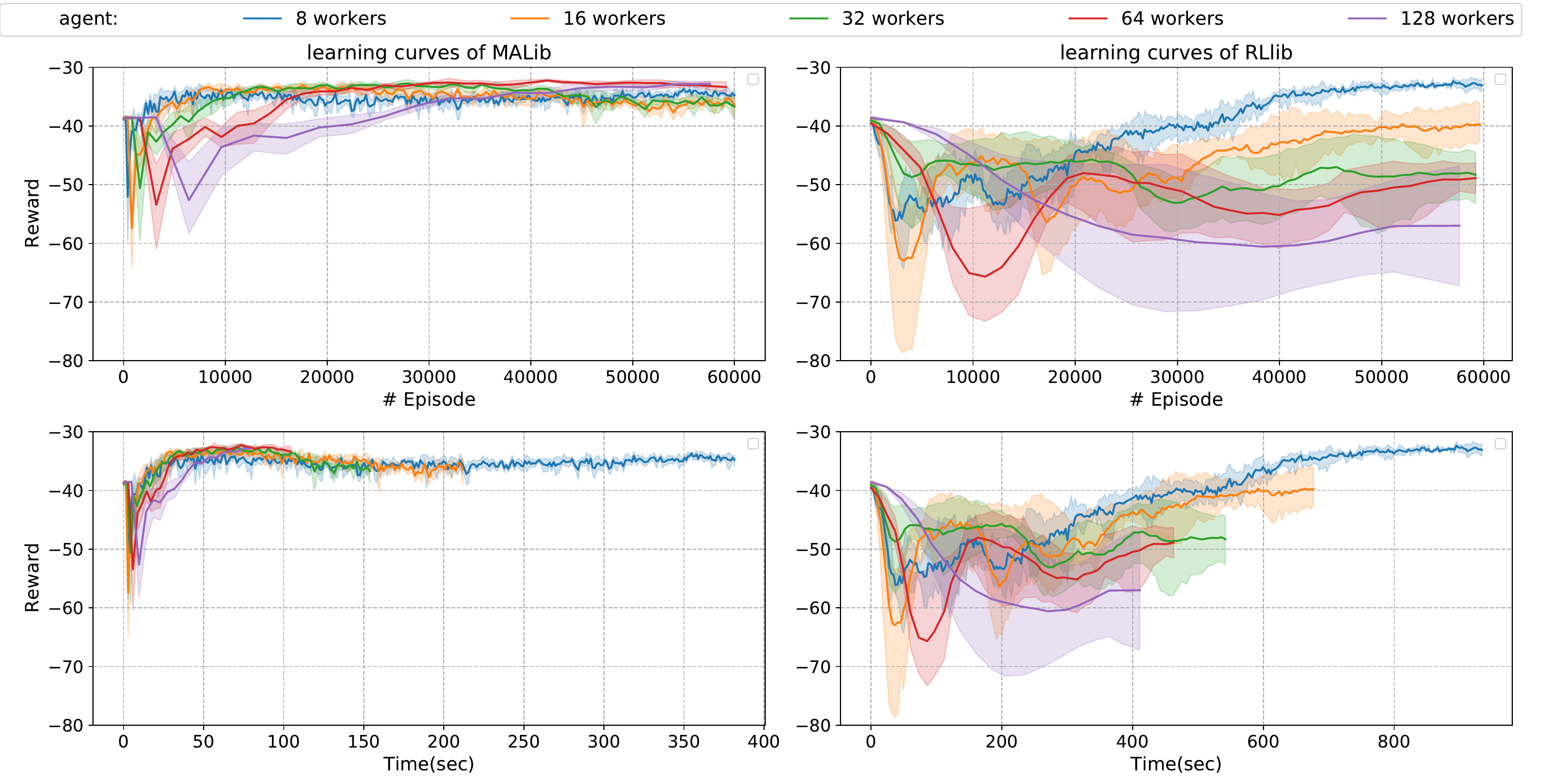}
    \caption{Comparisons between learning curves of MALib and RLlib when running MADDPG in \textit{simple spread} under different rollout worker settings.}
    \label{fig:mpe_spread}
\end{figure}

\paragraph{Simple Spread.} There are 3 agents, 3 landmarks in this scenario. Agents must learn to cover all the landmarks while avoiding collisions. More specifically, all agents are globally rewarded based on how far the closest agent is to each landmark (sum of the minimum distances). Locally, the agents are penalized if they collide with other agents (-1 for each collision). The relative weights of these rewards can be controlled with the local\_ratio parameter. Figure~\ref{fig:mpe_spread} shows the comparison from the converged reward and time-consumption.

\begin{figure}[ht!]
    \centering
    \includegraphics[scale=0.326]{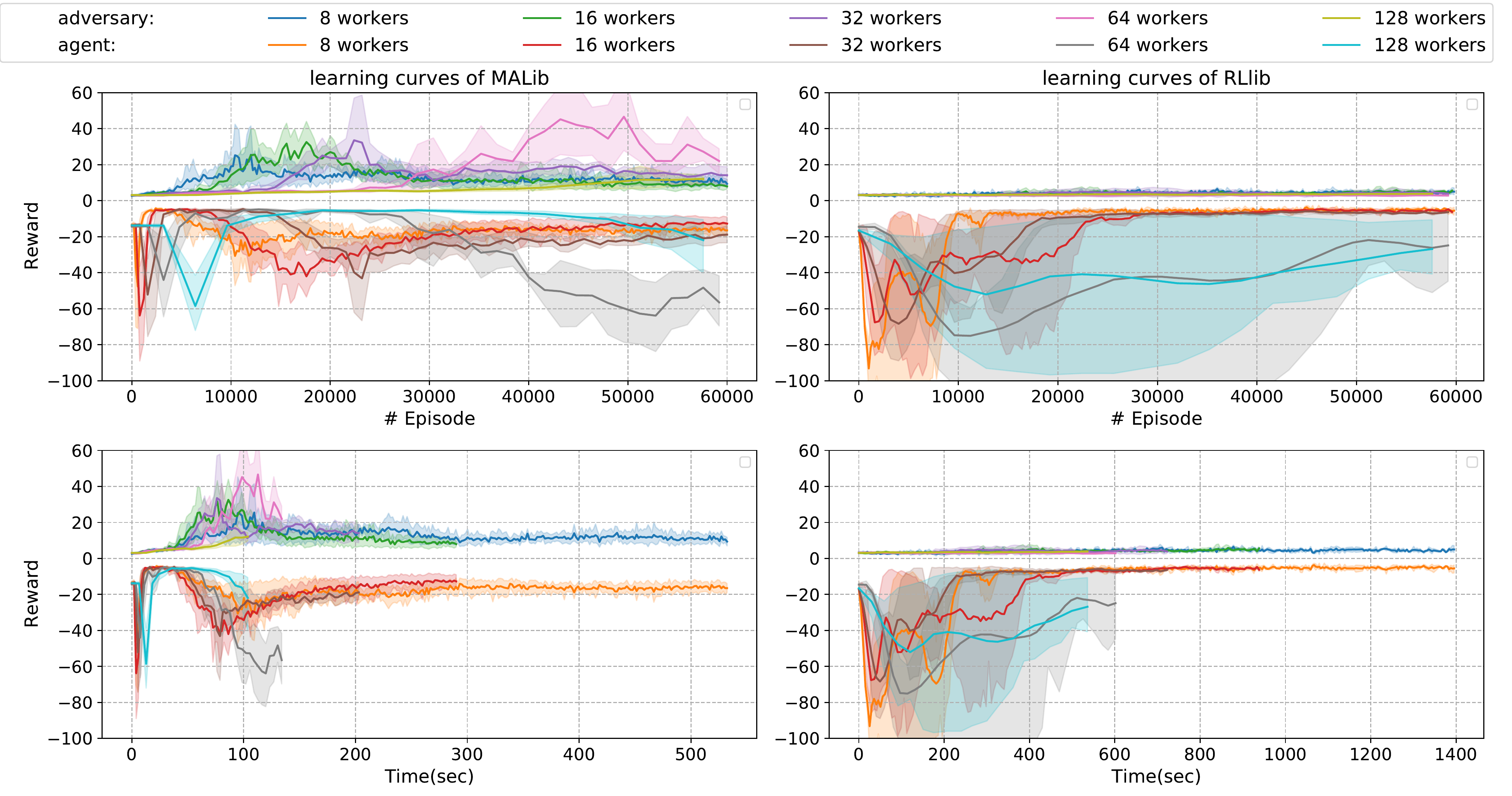}
    \caption{Comparisons between learning curves of MALib and RLlib when running MADDPG in \textit{simple tag} under different rollout worker settings.
    }
    \label{fig:mpe_tag}
\end{figure}

\paragraph{Simple Tag.} There are 1 good agent, 3 adversaries and 2 obstacles in this scenario. Good agent is faster and receive a negative reward for being hit by adversaries (-10 for each collision). Adversaries are slower and are rewarded for hitting good agents (+10 for each collision). Obstacles block the way. Figure~\ref{fig:mpe_tag} shows the comparison from the converged reward and time-consumption.

In summary, \framework{}'s implementation performs more steadily than RLlib in different worker settings, and achieved faster convergence rate.

\subsection{QMIX for SMAC}
\label{app:qmix_smac}
\textbf{S}tarCraft \textbf{M}ulti-\textbf{A}gent \textbf{C}hallenge~\cite{starcraft2} (SMAC) is a multi-agent benchmark that provides elements of partial observability, challenging dynamics, and high-dimensional observation spaces. It is built on top of a real-time strategic game, StarCraftII, creating a testbed for research in cooperative MARL tasks. We compared the \framework{}'s and PyMARL's implementations of QMIX. Table~\ref{table:smac_scenarios} shows the scenarios we tested, and the time consumption (seconds) when win rate achieves 80\%.

\begin{table}[ht!]
	\centering
	\renewcommand{\arraystretch}{1.}
	\captionsetup{type=table}
	\caption{Tested SMAC scenarios and time consumption when Win Rate = 80\%}
	\label{table:smac_scenarios}
	\resizebox{0.75\linewidth}{!}{
		\begin{tabular}{cccccc}
			\toprule
			\textbf{Scenario} & \textbf{Ally Units} & \textbf{Enemy Units} & \textbf{MALib} & \textbf{PyMARL} \\
			\midrule
			3m & 3 Marines & 3 Marines & \textbf{300} & 5625 \\
			8m & 8 Marines & 8 Marines & \textbf{500} & 12375  \\
			2s3z & 2 Stalkers \& 3 Zealots & 2 Stalkers \& 3 Zealots & \textbf{375} & 7920 \\
			3s5z & 3 Stalkers \& 5 Zealots & 3 Stalkers \& 5 Zealots & - & - \\
			\bottomrule
		\end{tabular}
	}
\end{table}

For the scenario \texttt{3s5z}, however, both of \framework{} and PyMARL cannot reach 80\% win rate.

\begin{figure}[ht!]
    \centering
    \includegraphics[width=\textwidth]{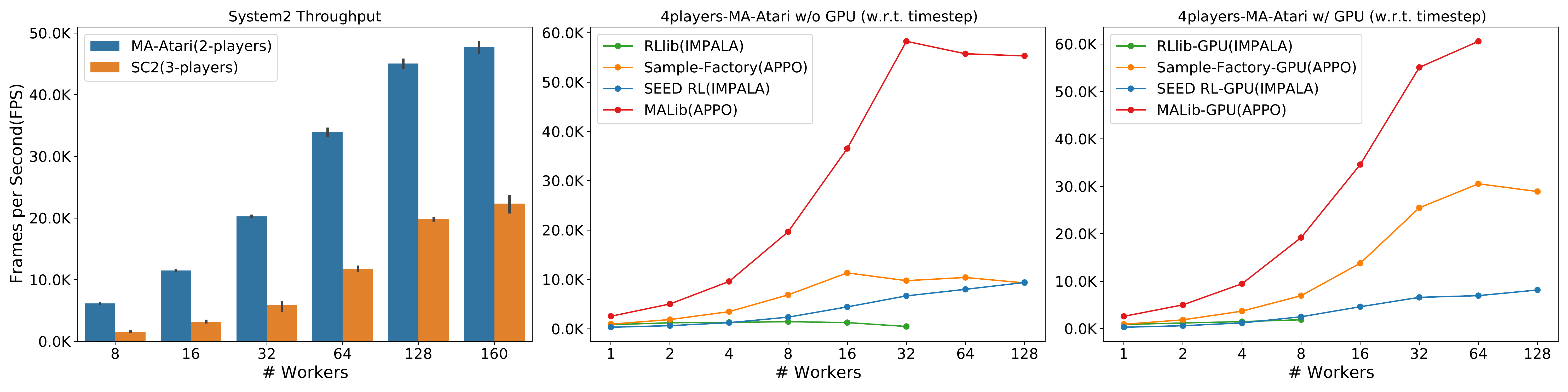}
    \vspace{-4mm}
    \caption{Additional comparisons between MALib and other distributed RL training frameworks. \textbf{(Left)}: System \#3 cluster throughput of MALib in 2-player MA-Atari and 3-player SC2. \textbf{(Middle)}: 4-player MA-Atari throughput comparison on System \#1 without GPU. \textbf{(Right)}  4-player MA-Atari throughput comparison on System \#1 with GPU.}
    \label{fig:thp-cmps}
\end{figure}

\subsection{Throughput and Comparisons}
In additional to the system $\#1,\#3$ descriptions included in the main body, the data throughput of \framework{} on system$\#$2 is depicted in Figure~\ref{fig:thp-cmps}~(Left) to show the performance of different system presets. Each group of experiments are repeated for five times and the corresponding throughput results are reported. As is shown, the effects of random seeds and fluctuations among different trails of throughput tests are insignificant, and therefore error bars are omitted in the following figures for simplicity. 

As illustrated in the Figure~\ref{fig:thp-cmps}~(Middle, Right), we further evaluated and compared the training throughput with three other distributed RL frameworks in a unequally configured multi-agent MA-Atari game(i.e. 4-player Pong-v1). 

Note that in the main text, we reported the results in a 2-player MA-Atari environment, which is less challenging than this case. And the FPS data obtained here has been re-aligned with timestamps for a fair comparison among all the frameworks, causing a degradation in the reported performance for Sample-Factory.

\section{User Examples}

\subsection{Parallel Training Customization}
\framework{} provides four basic training interfaces to support MARL training in different paradigms and distributed computing settings. We present an use case of centralized training customization to show the convenience of training customization. Specifically, the only thing we need to do is to customize a Trainer class which inherits from a standard training interface \texttt{Trainer}. Other training customization is similar the this case.

\begin{verbatim}
# ==== Centralized training customization ====
from malib.agents import CenntralizedAgent
from malib.algorithms.common import Trainer
from malib import runner

# ==== Customized trainer ====
class CentralizedTrainer(Trainer):
    # centralized trainer implemented the logic 
    # of centralized training for arbitrary 
    # independent RL algorithm like DDPG, A2C or PPO.
    def optimize(self, batch):
        pass
        
    def preprocess(self, batch, **kwargs):
        # merge sample batch or implement communication
        # proxy here
        pass

runner.run(
    # here, we let all agents share the same 
    # centralized training interface
    agent_mapping_func=lambda agent: "share",
    # set
    training={
        "interface": {
            "type": CentralizedAgent,
            # ...
        }
    },
    # specify the trainable RL algorithm and trainer
    algorithms={
        "custom_name": {
            "PPO": {
                "trainer": CentralizedTrainer
                # ...
            }
        }
    },
    # ...
)
\end{verbatim}

\label{app:hyper}

\subsection{Single Instance Examples}
Though \framework{} is designed to support distributed PB-MARL training tasks, it also supports single instance training via decoupled rollout and training interfaces. We present a single instance of PSRO training here.

\paragraph{Configuration Initialization.} At the first, we need to initialize the training and rollout congurations, also the interface instances.

\begin{verbatim}
exp_cfg = {
    "expr_group": "single_instance_psro",
    "expr_name": f"simple_tag_{time.time()}",
}
env_config = {
    "num_good": 1,
    "num_adversaries": 1,
    "num_obstacles": 2,
    "max_cycles": 75,
}

env = simple_tag_v2.env(**env_config)
possible_agents = env.possible_agents
observation_spaces = env.observation_spaces
action_spaces = env.action_spaces

env_desc = {
    "creator": simple_tag_v2.env,
    "config": env_config,
    "id": "simple_tag_v2",
    "possible_agents": possible_agents,
}

# agent buffer, for sampling
agent_episodes = {
    agent: Episode(
        env_desc["id"],
        policy_id=None, 
        capacity=args.buffer_size
    )
    for agent in env.possible_agents
}

rollout_config = {
    "stopper": "simple_rollout",
    "metric_type": args.rollout_metric,
    "fragment_length": 100,
    "num_episodes": 1,
    "terminate": "any",
    "mode": "on_policy",
    "callback": rollout_wrapper(agent_episodes),  # online mode
}

algorithm = {
    "name": args.algorithm, 
    "model_config": {},
    "custom_config": {}
}

learners = {}
for agent in env.possible_agents:
    learners[agent] = IndependentAgent(
        assign_id=agent,
        env_desc=env_desc,
        training_agent_mapping=None,
        algorithm_candidates={args.algorithm: algorithm},
        observation_spaces=observation_spaces,
        action_spaces=action_spaces,
        exp_cfg=exp_cfg,
    )
    learners[agent].register_env_agent(agent)

rollout_handlers = {
    agent: RolloutWorker(
        worker_index=None,
        env_desc=env_desc,
        metric_type=args.rollout_metric,
        remote=False,
        exp_cfg=exp_cfg,
    )
    for agent in env.possible_agents
}

# global evaluator to control the learning
psro_evaluator = PSROEvaluator()

# payoff manager, maintain agent payoffs and simulation status
payoff_manager = PayoffManager(
    env.possible_agents,
    exp_cfg=exp_cfg
)
\end{verbatim}

\paragraph{Training Workflow.} The training workflow is comprised of rollout and policy optimization. The implementation is listed as follows.

\begin{verbatim}
def training_workflow(
    trainable_policy_mapping: Dict[AgentID, PolicyID]
):
    # ...
    for agent in env.possible_agents:
        policy_distribution = payoff_manager.get_equilibrium(
            population)
        policy_distribution[agent] = {
            trainable_policy_mapping[agent]: 1.0}
        rollout_handlers[agent].ready_for_sample(
            policy_distribution)

        for epoch in range(args.num_epoch):
            # ==== collect training data ====
            rollout_feedback[agent], _ = \
                rollout_handlers[agent].sample(
                    callback=rollout_config["callback"],
                    num_episodes=[rollout_config["num_episodes"]],
                    threaded=False,
                    role="rollout",
                    trainable_pairs=trainable_policy_mapping,
                )
            
            # ==== policy optimization ====
            batch = agent_episodes[agent].sample(args.batch_size)
            res = learners[agent].optimize(
                policy_ids={
                    agent: trainable_policy_mapping[agent]
                },
                batch={agent: batch},
                training_config={
                    "optimizer": "Adam",
                    "critic_lr": 1e-4,
                    "actor_lr": 1e-4,
                    "lr": 1e-4,
                    "update_interval": 5,
                    "cliprange": 0.2,
                    "entropy_coef": 0.001,
                    "value_coef": 0.5,
                },
            )
            training_feedback.update(res)

    return {
        "rollout": rollout_feedback,
        "training": training_feedback
    }
\end{verbatim}

\paragraph{Main Loop.}
\begin{verbatim}
# init agent with fixed policy
policy_mapping = extend_policy_pool(trainable=True)
equilibrium: Dict[
    AgentID, Dict[PolicyID, float]
] = run_simulation_and_update_payoff(policy_mapping)

# === Main Loop ===#
iteration = 0
while True:
    # 1. add new trainable policy
    trainable_policy_mapping = extend_policy_pool(trainable=True)

    # 2. do rollout and training workflow
    feedback: Dict = training_workflow(trainable_policy_mapping)

    # 3. simulation and payoff table update
    equilibrium: Dict[
        AgentID, Dict[PolicyID, float]
    ] = run_simulation_and_update_payoff(trainable_policy_mapping)

    # 4. convergence judgement
    nash_payoffs: Dict[AgentID, float] = payoff_manager.aggregate(
        equilibrium=equilibrium
    )
    weighted_payoffs: Dict[AgentID, float] = \
        payoff_manager.aggregate(
            equilibrium=equilibrium,
            brs=trainable_policy_mapping,
        )
    evaluation_results = psro_evaluator.evaluate(
        None,
        weighted_payoffs=weighted_payoffs,
        oracle_payoffs=nash_payoffs,
        trainable_mapping=trainable_policy_mapping,
    )

    if evaluation_results[EvaluateResult.CONVERGED]:
        print("converged!")
        break

    iteration += 1
\end{verbatim}

\end{document}